\documentstyle[12pt,epsfig]{article}

 1
\def\as{\alpha_{\rm S}}

\def\citenum#1{{\def\@cite##1##2{##1}\cite{#1}}}
\def\citea#1{\@cite{#1}{}}

\def\as{\alpha_{\rm S}}

\def\g{\gamma}

\def\s{\sigma}

\def\({\left(}
\def\){\right)}

\def\citenum#1{{\def\@cite##1##2{##1}\cite{#1}}}
\def\citea#1{\@cite{#1}{}}

\def\l1vt{\vec{l_{1\perp}}}

\def\rt{r_{\perp}}
\def\bt{b_{\perp}}
\def\rt2{$r^2_{\perp}$}
\def\bt2{$b^2_t$}

\def\jol1{$J_0(\,l_{1\perp}\,r_{\perp}\,)$}

\def\citea#1{\@cite{#1}{}}

\def\beq{\begin{equation}}
\def\eeq{\end{equation}}
\def\bea{\begin{eqnarray}}
\def\eea{\end{eqnarray}}

\def\eq#1{{Eq.~(\ref{#1})}}

\def\bbbz{{\mathchoice {\hbox{$\sf\textstyle Z\kern-0.4em Z$}}
{\hbox{$\sf\textstyle Z\kern-0.4em Z$}}
{\hbox{$\sf\scriptstyle Z\kern-0.3em Z$}}
{\hbox{$\sf\scriptscriptstyle Z\kern-0.2em Z$}}}}
\def\npb#1#2#3{    {\it Nucl. Phys. }{\bf B#1} (19#2) #3}
\def\plb#1#2#3{    {\it Phys. Lett. }{\bf B#1} (19#2) #3}
\def\prd#1#2#3{    {\it Phys. Rev. }{\bf D#1} (19#2) #3}

\def\zpc#1#2#3{    {\it Z. Phys. }{\bf C#1} (19#2) #3}

\def\sjnp#1#2#3{   {\it Sov. J. Nucl. Phys. }{\bf #1} (19#2) #3}

\def\l{\lambda}

\begin{document}

\begin{titlepage}
\begin{flushright}
TAUP\,\,2443 - 97\\
DESY \,\,97 - 154\\
July 1997\\
{\bf hep-ph/9708275}
\end{flushright}

\begin{center}
{\Large\bf{A UNITARITY BOUND AND THE COMPONENTS }}\\[1.5ex]
{\Large \bf{OF  PHOTON - PROTON INTERACTIONS}}\\[6ex]

{\large \bf { E. Gotsman ${}^{a)\,b)}$${}^*$\footnotetext{ ${}^*$ E-mail:
gotsman@post.tau.ac.il}
,
 E.M. Levin ${}^{a)\,b)\,c)}$$^{**}$\footnotetext{${}^{**}$
E-mail:leving@post.tau.ac.il}
  and 
 U. Maor ${}^{a)\dagger}$
\footnotetext{$^{\dagger}$ E-mail: maor@post.tau.ac.il}}} \\[2.5ex]

{\it ${}^{a)}$ School of Physics and Astronomy, Tel Aviv University}\\
{\it Ramat Aviv, 69978, ISRAEL}\\[1.5ex]
{\it ${}^{b)}$ DESY Theory, Notkestr. 85,D-22603, Hamburg,
GERMANY}\\[1.5ex]
{\it$ {}^{c)}$ Theory Department, Petersburg Nuclear Physics Institute}\\
{\it 188350, Gatchina, St. Petersburg, RUSSIA}\\[8.5ex]
\end{center}

{\large \bf Abstract:}
We show how and why the short  distance (``hard")
interaction, which is calculated  in perturbative QCD, provides a mass
cutoff in
  Gribov's formula for photon-proton collisions. This enables us
to find   a new  and 
more restrictive unitarity bound for this process,
$\sigma(\gamma^{*}p)\leq C(ln\frac{1}{x})^{\frac{5}{2}}$.
We develop 
 a simple model
 that consists of ``soft" and ``hard" contributions, which yields a
qualitative
 description of
  the published  experimental data over  a wide range  of
 photon virtualities ($Q^2$) and energies ($W$). This model 
provides  a
quantitative way of evaluating  the relative rate of the short  and long 
distance
contributions, in the 
different kinematic regions. The main results of the analysis are  (i)
that even at $Q^2 =0$ and  high energies the 
short  distance contribution is not small,
 and it provides a possible explanation for  the experimental rise of the
 photoproduction cross section; and (ii) at large values of $Q^2$,  the 
long  distance processes  still contribute to the
total cross section.

\end{titlepage}

\section{Introduction.}
The total cross section of a hadron-hadron interaction is bound by
the Froissart - Martin limit \cite{FRST}
\beq \label{FRST}
\s_{tot}\,\,\leq\,C\,\ln^2\frac{s}{s_0}\,\,,
\eeq
where $C\,=\,\frac{\pi}{\mu^2}$,  depends on the mass of the lightest
particle exchanged in the crossed channel. The bound is a consequence of
 $s$-channel unitarity, analyticity and crossing symmetry. In spite  of the 
ambiguity 
in the determination of $C$, we suggest that at the  presently available
hadron accelerator energies one should look  for
 phenomena  associated with $s$ - channel unitarity,  rather than the
absolute bound. Indeed, a
careful study \cite{GLM}  shows that  the scale of saturation
of $s$-channel
unitarity
 in elastic $\bar p p$ reactions is above the Tevatron
energy, while  the saturation scale   for diffractive channels is
considerably
lower, appearing at ISR energies. This qualitative theoretical study is
strongly supported by the experimental observation that, whereas
$\s_{el}/\s_{tot}$ grows all through the ISR - Tevatron range, the ratio
$\s_{diff}/\s_{tot}$  decreases with energy \cite{TEVATDD}.

The study of unitarity and the Froissart - Martin bounds in DIS are more
complicated than for the  hadron - hadron case.  These complications
originate from ambiguities in the implementation of the unitarity
constraints due to electromagnetic photon coupling and the absence of a
proper elastic channel, as well as the introduction of the mass of the
virtual photon as an additional kinematic variable.
From
a phenomenological point of view, we have to take into account the
 ``soft" and ``hard", or alternatively, the long distance and 
short distance phenomena, as contributors to the total $\g^* p$ cross
section, or the target structure function. This 
differs from the  usual picture  of the hadron - hadron collision,
where the incoming time like particle masses are fixed, and the total cross
section is determined by the ``soft" ( long distance ) Pomeron.

A remarkable simplification of the DIS analysis has been suggested by
Gribov \cite{GRBV} in the context of DIS on a nuclear target.
 Gribov's
main observation was that at
high energies, the $\g^*$ fluctuates into a hadronic system ( i.e. $\bar q
q
$ to the lowest order ) with a coherence length, \,$l_c\,=\,\frac{1}{mx}$,
which is much  larger than the target radius.
 $m$ denotes  the target mass and $x$  the
Bjorken
scaling variable ( $x\,=\,\frac{Q^2}{s}$, where $Q^2$ is the photon
virtuality ). Hence, we can  describe DIS as a two step process

1) The $\g^*$ transforms into a hadronic system well before the
interaction with the target.

2) The produced hadronic system interacts with the target.

Gribov added two technical assumptions, which simplified the calculation

a) The   hadronic interaction is a black disc interaction. This
assumption was made for a heavy nuclear target.  
  In the black disc limit, the strong interaction diffractive
dissociation channels $M^2 p \,\rightarrow \,M'^2 p$   with
$M\,\neq\,M'$
can be neglected.

b) A dispersion relation,  without subtractions, can be written in terms
of 
the variable $M^2$.

The resulting DIS cross section is then written  as
\beq \label{GRM}
\s(\g^* N )\,\,=\,\,\frac{\alpha_{em}}{3 \,\pi}\,\int\frac{R(M^2)\,M^2\,d
\,M^2}{(\, Q^2\,+\,M^2\,)^2}\,\s_{M^2N}(s)\,\,.
\eeq
Here $R(M^2)$  is defined as the ratio
\beq \label{RATIO}
R(M^2)\,\,=\,\,\frac{\s(e^{+} e^{-} \,\rightarrow\,hadrons)}{\s(e^{+}
e^{-}\,\rightarrow\,\mu^{+} \mu^{-} )}\,\,.
\eeq
The  notation is illustrated  in Fig.1 where $M^2$ is the mass squared of
the
scattered hadronic system, $\Gamma^2(M^2)\,=\,R(M^2)$  and  $\s_{M^2 
N}(s)$
 is the cross section for the hadronic system to scatter off the 
nucleonic target.

\begin{figure}
\centerline{\psfig{file= 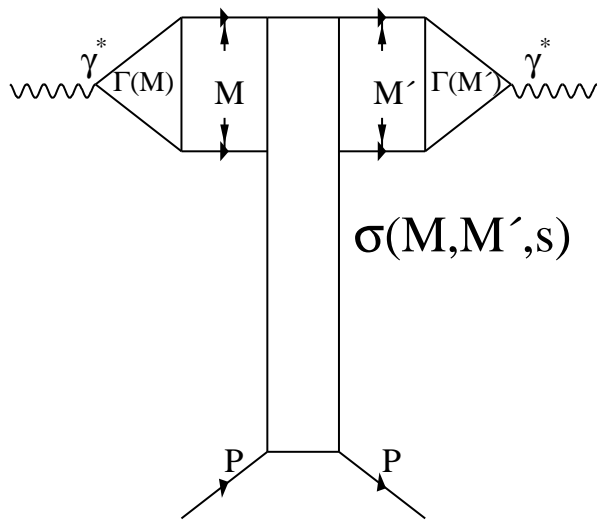,width=100mm}}
\caption{The generalized Gribov's formula for DIS.}
 \end{figure}
 Assuming \eq{FRST} for $\s_{M^2 N}$ and integrating \eq{GRM} over  $M^2$
we obtain      a  $\gamma^{*} p$ cross section  which behaves like $
\ln(\,\frac{M^2_{max}\,+\,Q^2}{M^2_{min}\,+\,Q^2}\,)$. We note that
$M^2_{min} \,=\,4m^2_{\pi}$, whereas $M^2_{max}\,\propto\,s$, therefore, 
one
 easily obtains  the  Gribov's bound \cite{AFS}
 \beq \label{GFB}
\s(\g^* N )\,\,\leq\,\,\frac{\alpha_{em}}{3 \,\pi} R_{\infty} 
C ln^2\frac{s}{s_0}\,\ln\frac{1}{x}\,\,,
\eeq
here $R_{\infty}$ denotes the ratio given by \eq{RATIO} in the  high 
energy
limit. The logarithmic behaviour of the above bound is unchanged by the
introduction of an arbitrary  
high mass  cutoff   $\frac{M^2}{s}\,\leq\, 0.1$    \cite{AFS}.
A disturbing feature of \eq{GFB}, is that it is less stringent than the
Froissart - Martin bound, and we question whether this is a genuine
feature
of DIS, or an artifact of our approach and the assumed  input.  In
particular,
an extra logarithmic power of $M^2$ which  appears in \eq{GRM}  due to the 
upper limit of the integration,   does not appear explicitly  in
Gribov's
formula.

Attempting to clarify these problems, we  present  a calculation
that maintains Gribov's basic hypothesis, which we find attractive , but
gives up the black disc assumption for large values of $M^2$.
  The physical reason why the black disc assumption cannot be correct,
even for a very heavy nucleus, or at extremely high energies in the
Froissart
- Martin region, is simple. 
 The quark - antiquark pair with
 a large mass has a small  transverse size, typically  of the order of
$r_{\perp}\,\propto\,\frac{1}{M}$. Being colour neutral such a pair
can penetrate  without interacting  through
 a large target such as a heavy nucleus, or a hadron at ultra high
energies.
The attractive  feature of this scenario is  that at large
 $M^{2}$ the  interaction takes place
 at small distances, or in other words, it is a hard process which
can be calculated in the framework of perturbative QCD (pQCD).

 In section
2 we  develop a general method  to take into account the effect described
above,  and show that \eq{GFB}   should be replaced by
the following relation:
\beq \label{CORGB}
\s(\g^* N )\,\,\leq\,\,\frac{\alpha_{em}}{3 \,\pi}  R_{\infty} 
C  \ln^2\frac{s}{s_0}\,\ln (\,\frac{Q^2\, +\,M^2(x)}{Q^2
\,+\,M^2_0}\,)\,\,,
\eeq
where $M^2(x)$ is the solution of the equation
\beq \label{EQMX}
\frac{4 \pi \as}{3\, R^2_N\, M^2(x)}
\,\,xG^{DGLAP}(x,M^2(x))\,\,=\,\,1\,\,.
\eeq
 $xG$ denotes  the gluon density of  the target, and $R^2_N$  the gluon
correlation radius, that has been estimated using the HERA  diffractive
dissociation data in Ref.\cite{GLMSLOPE}. $M^2_0$ is a cutoff in  mass
that separates the ``soft" ( lomg  distance ) processes from the ``hard"
( short  distance ) ones. Clearly, the assumption that the production
for  $M^{2} < M^{2}_{0}$ is soft, whereas $M^{2} \geq M^{2}_{0}$ is hard,
is an  oversimplification.
 In  our approach, the value of
$M^2_0$ is a pure phenomenological parameter, which we  determine  from
a fit to
 the experimental DIS data at sufficiently small values of $Q^2$.

In section 3 we develop a phenomenological approach based on our general
formulae of section 2, that allows us to match  the DIS and real
photoproduction data. The main idea underlying our approach is to 
parameterize the low
mass region using  Gribov's formula for the ``soft" processes,
while for the
high $M^2$ contribution, the leading $\as \ln(1/x)$ approximation of pQCD
is  used. We show that with the choice $M^2_0\, \approx\, 5\,GeV^2$, we are
able to qualitatively  describe the main features of the experimental data
for photon
nucleon interactions at high energy, and for most  values of the photon
virtuality ( $Q^2$ ).
\section{General formalism and a Gribov - Froissart      type \\
 bound for DIS}
\subsection{A generalization of the Gribov's formula}
As  mentioned above , Gribov argued that  one  can use a 
 dispersion relation with respect to the masses $M$ and $M'$ 
 to describe  the photon - hadron interaction  ( see Fig.1 for
 notation ), as  the  correlation length $l_c =
\frac{1}{m x}\,\gg\,R_N$,  the target size. Based on this idea
we can write a  general formula for the  photon - hadron interaction,

\beq \label{GF}  
\s(\g^* N )\,\,=\,\,
\frac{\alpha_{em}}{3\,\pi}\,\int \frac{\Gamma(M^2)\,\,d\,M^2}{\,
Q^2\,+\,M^2\,}
\,\,\s(M^2,M'^2,s)\,\,
\frac{\Gamma(M'^2)\,\,d\,M'^2}{\, Q^2\,+\,M'^2\,} \,\,.
\eeq
 In the black disc approximation $
\s(M^2,M'^2,s)\,=\,2\,\pi \,R^2_N\,M^2\,$\,\,$\delta(M^2\,-\,M'^2)$,
which
leads
to
the Gribov's formula of \eq{GRM}.  \eq{GF} enables us  to separate the
``soft" 
 and ``hard" interactions,
by  introducing a separation scale $M_0$ in the integral over the  masses
$M$ and $M'$ in \eq{GF}.  \eq{GF} can  be
rewritten in the form
\beq \label{SUMSH}
 \s(\g^* N)\,\,=\,\,\s^{soft}\,\,+\,\,\s^{hard}\,\,,
\eeq
where
\beq \label{SOFT}
\s^{soft}\,\,=\,\,\frac{\alpha_{em}}{3\,\pi}\,\int^{M^2_0}_{4m_{\pi}^{2}}
\frac{\Gamma(M^2)\,\,d\,M^2}{\,Q^2\,+\,M^2\,}
\,\,\s(M^2,M'^2,s)\,\,
\frac{\Gamma(M'^2)\,\,d\,M'^2}{\, Q^2\,+\,M'^2\,}\,
\eeq
$$
\leq\,
\frac{\alpha_{em}}{3 \,\pi}\,\int^{M^2_0}_{4m_{\pi}^{2}}
\frac{R(M^2)\,M^2\,d
\,M^2}{(\, Q^2\,+\,M^2\,)^2}\,\,\s_{M^2\,N}(s)\,\,.
$$
Here, we have used  Gribov's ideas to estimate the contribution of
the ``soft" processes
 using the black disc approximation.  It would be preferable if
 the
``soft" contribution could  be calculated in 
nonperturbative QCD (npQCD ). Thus far,
unfortunately,  no consistent npQCD approach has
been
developed   for this contribution and what we have
at  hand are  only  phenomenological parameterizations   describing the
``soft"
hadron processes.  We observe that experimentally
$\frac{\s_{diff}}{\s_{el}}$ is decreasing rapidly with energy    in
nucleon
- nucleon interactions \cite{ TEVATDD}.
In our context this translates to the observation that a transition
from
 a  hadronic  system with mass $M$ to one with mass $M'$ is
somewhat smaller than
the
elastic cross section which does not change the value of the mass.
Moreover, we expect theoretically the ratio $\frac{\s_{diff}}{\s_{el}}$
to decrease logarithmically in the high energy limit \cite{GLM}.
 These   observations support   our idea that a  suppression  of
$M$ to
$M'$
transition can be used as  the first order estimate of the ``soft" 
 contribution in \eq{GF}.
Clearly, at this stage,   any description of the ``soft" term
in \eq{SUMSH} has to be based on a  model. 

An  attractive feature of  our approach
 is the  introduction    of  the separation scale
$M_0$. This  allows us to use both   the ``soft" high energy
phenomenology as well as the pQCD calculation for the photon -
hadron interaction at high energy.

Accordingly, for the short  distance  contributions we have 
\beq  \label{HARD}
\s^{hard}\,\,=\,\,\frac{\alpha_{em}}{3\,\pi}\,\int^{\infty}_{M^2_0} 
\frac{\Gamma(M^2)\,\,d\,M^2}{\,
Q^2\,+\,M^2\,}
\,\,\s^{QCD}(M^2,M'^2,s)\,\,
\frac{\Gamma(M'^2)\,\,d\,M'^2}{\, Q^2\,+\,M'^2\,} \,\,,
\eeq
where we can  use the leading log(1/x) approximation of pQCD to evaluate
this integral.

\subsection{The ``hard" contribution to the generalized Gribov's formula}
 We wish to  rewrite the  formula for the ``hard" DIS cross section
in a  form which is similar  to  \eq{HARD}.
 The cross section for  DIS in the region of
small $x$ ( high energy ) in the  leading log($1/x$) approximation of pQCD,
has the form \cite{LR87} \cite{MU90}
 
\beq \label{DISCRSEC}
\s^{QCD}_{\g^* p}\,\,=\,\,\int^1_0 d z\, \int \,d^2 r_{\perp}
\,\,|\Psi^{\g^*}(Q^2,z,r_{\perp})|^2 \,\,\,\int d^2 b_t
\s_N(x,r_{\perp},b_t)\,\,,
\eeq
where $\Psi^{\g^*}$ is the wave function of the virtual photon.
Although  the separation   between the ``soft" and ``hard" sectors is
more natural in the analysis of longitudinal polarized photons
\footnote{This subject will be further discussed in section 3.}, we limit
our discussion at this stage to 
transverse polarized photons as this  gives  the dominant contribution to  
the total cross
section. The calculations pertaining to the   longitudinal polarized
photons will be
published elsewhere.

  For
a transverse polarized photon we have \cite{NNN} \beq \label{SQWF}
| \Psi^{\g^*}_T |^2\,\,=\,\,\frac{\alpha_{em} N_c}{2 \pi^2}\,\sum_{f}
\,Z^2_f \,[ z^2 + ( 1 - z )^2 ]\, {\bar Q}^2\,K_1^2(\bar Q\,r_{\perp})\,\,,
\eeq
where $K_1$ is the modified Bessel function, ${\bar Q}^2\,=\,Q^2 z (1 -
z)$, and $N_c$  the number of colours.
 $Z_f$ and  $z$  are  the fraction of the charge, and the
fraction of energy carried by the quark.
$r_{\perp}$  denotes the transverse splitting
between the quark and antiquark.  $\s_N(x,r_{\perp},b_t)$ is the cross
section of the colour dipole of a  size $r_{\perp}$ with the target
at fixed impact parameter $b_t$. This 
cross section is equal to
\beq \label{SB}
\s_N(x,r_{\perp},b_t)\,\,=\,\,2\,Im\,a_{el} (x,r_{\perp},b_t)\,\,,
\eeq
where $a_{el}$ is the elastic amplitude in the $b_t$ - representation,
which is closely related to  the scattering amplitude   of the dipole at
a definite value of the transfer momentum  squared $t \,=\,-\,q^2_t$
\beq \label{SBQ}
a_{el}(x,r_{\perp},b_t)\,\,=\,\,\frac{1}{2 \pi}\,\int\,d^2 q_t \,e^{ - i
\vec{q}_t\cdot \vec{b}_t}\,A(x,r_{\perp},t)\,\,.
\eeq
 Since high energy experimental data suggest that $Re\, 
a_{el}\,\,\ll\,\,Im\,a_{el}$,
 $s$-channel unitarity  implies  \cite{LR87}\cite{MU90}
\beq \label{CRSECUN}
\s_N(x,r_{\perp},b_t)\,\,=\,\,2\,\{\,\,1\,\,-\,\,e^{- \frac{1}{2}
\,\Omega}\,\,\}\,\,,
\eeq 
with arbitrary real function  $\Omega$.  In the kinematic region where
$\Omega\,\ll\,1$, for dipoles of  small sizes $ r_{\perp} \,\ll R_N$,
this function  is equal to  \cite{AGLFRST}
\beq \label{OMEGA}
\Omega\,\,=\,\,S(b_t)\,\frac{ \pi^2 \as}{3}\,r^2_{\perp}\,x
G(x,\frac{4}{r^2_{\perp}})\,\,.
\eeq
 $S(b_t)$ is the nonperturbative
two gluon form factor, which normalizes as $ \int d^2 b_t  S(b_t) \,=\,1$. 
 From general principles of analyticity we only know its large $b_t$
behaviour,
$S(b_t)\,|_{b_t\,\rightarrow\,\infty} \,\,\rightarrow\,\,e^{-\,2\,\mu\,
b_t}$, where $\mu$ is the mass of the lightest hadron ( pion ).

Many practical applications  assume an  exponential
parameterization for $S(b_t)$ of  the form
\beq \label{BFF}
S(b_t)\,\,=\,\,\frac{1}{\pi R^2_N}\,\,e^{- \frac{b^2_t}{R^2_N}}\,\,,
\eeq
where $R^2_N$ is the correlation length between two gluons in the proton.
For the case of uncorrelated gluons $R_N$ is the hadron ( proton ) radius.
 
  We wish  to comment on the form of \eq{OMEGA}, recalling  the  standard
procedure for
 solving  the DGLAP evolution equations.

 1) The first step: we introduce
the moments of the parton density, 
$$
x G(x,Q^2)\,=\,\frac{1}{2\pi i}\int_C e^{-\omega \,\ln(1/x)} \,M(\omega,
Q^2) \,d \omega,
$$
 where the
contour
$C$ is located to the right of all the singularities of the moment
$M(\omega,
Q^2)$.

 2)   The second step: we find the solution to the DGLAP equations
for the moment
\beq \label{OMEGA1}
\frac{d M(\omega, Q^2)}{d \ln Q^2}\,\,=\,\,\gamma(\omega)\,M(\omega,
Q^2)\,\,.
\eeq  
The solution is
\beq \label{OMEGA2}
M(\omega, Q^2)\,\,=\,\,M(\omega,Q^2_0)\,
\,e^{\gamma(\omega)\,\ln(Q^2/Q^2_0)}\,\,.
\eeq
Here $M(\omega,Q^2_0)$ is the nonperturbative input which is  taken either
from experimental data or from the  ``soft" ( model dependent ) 
phenomenology .   

 3)  The third step: we find the solution for the parton density
using the inverse transform
\beq \label{OMEGA4} 
xG(x,Q^2) \,\,=\,\,\int_C \,\frac{d \omega}{2 \pi i}\,\,e^{\omega
\,\ln(1/x)\,+\,\gamma(\omega)\,\ln(Q^2/Q^2_0)}\,M(\omega,Q^2_0)\,\,.
\eeq

We conclude that in order  to find a solution of the DGLAP equation we
need to know the
nonperturbative input $M(\omega,Q^2_0)$ and the anomalous dimension
$\gamma(\omega)$, which we can calculate in pQCD.  To obtain
the $b_t$ - dependence of the deep inelastic structure function we have to
calculate the $t$ - dependence of the imaginary part of the virtual
photon Compton amplitude ( see \eq{SB} and \eq{SBQ} ). In the  framework
of the DGLAP evolution equations  we
have two different regions of $t$ :  (i)  $t\,\leq \,Q^2_0$ and (ii)
$t\,\geq\,Q^2_0$. For the case when  $t\,\geq\,Q^2_0$,   $t$
defines the
factorization scale and replaces $Q^2_0$ in \eq{OMEGA4} ( see
Ref.\cite{GLR} ). However, for $t\,\leq\,Q^2$ the factorization scale is
equal to $Q^2_0$ and the only $t$ - dependence is  concentrated in
$M(\omega,Q^2_0;t )$. The factorizable form of the
initial moments  $M(\omega,Q^2_0;t)$\,=\,$M(\omega,Q^2_0)\, 
F(t) $ is certainly an assumed  model, but this model is reasonable for
$t\,\ll
\,Q^2_0$. It should be stressed that this assumption  which led to  the
explicit  form
of
\eq{OMEGA}, is not needed for the large $b_t$ behaviour, which
is the only
ingredient  of \eq{OMEGA} used for the proof of the Gribov - Froissart
bound for DIS.

 Using \eq{CRSECUN}, we can distingwish between two kinematic limits that
we use for our approximation.

I)\,   $\Omega\,\ll\,1$ and
$  \s_N(x,r_{\perp},b_t)\,\rightarrow\,\Omega$\,\,
with $\Omega$ given by \eq{OMEGA}.

II)\,  $\Omega\,\gg\,1$  where
$ \s_N(x,r_{\perp},b_t)\,=\,2\,\{\,1\,-\,e^{-
\frac{1}{2}\Omega}\,\}\,
\rightarrow\,2 \,\,.$

~

At each fixed $x$ and $r_{\perp}$ the boundary between these two regions
occurs  at $b_t = b_0$, which  can be determined from the equation

\beq \label{B0}
S(b_0)\,\frac{ \pi^2 \as}{3}\,r^2_{\perp}\,x
G(x,\frac{4}{r^2_{\perp}})\,\,=\,\,1\,\,.
\eeq
Substituting the large $b_t$ behaviour of the form factor $S(b_t)$, one 
finds
\beq \label{FB0}
b_0\,\,=\,\,\frac{1}{2\mu}\,\,\ln[r^2_{\perp}xG(x,\frac{4}{r^2_{\perp}})]\,\,.
\eeq

 In the kinematic region II we  rewrite \eq{DISCRSEC} in the form of
\eq{HARD}. This  is a very simple task once  we recall that 
\beq \label{MCD}
\bar Q K_1( \bar Q r_{\perp} )\,\,=\,\,\int\,\, \frac{k^2\,d k}{{\bar
Q}^2\,+\,k^2}\,\,J_0(k r_{\perp})\,\,,
\eeq
or
\beq \label{MCDSQ}
[\,\bar Q K_1( \bar Q r_{\perp} )\,]^2\,\,=\,\,\int\,\, \frac{k^2_1\,d
k_1}{{\bar
Q}^2\,+\,k^2_1}\,\,J_0(k_1 r_{\perp})\,\,\int\,\, \frac{k^2_2\,d
k_2}{{\bar
Q}^2\,+\,k^2_2}\,\,J_0(k_2 r_{\perp})\,\,.
\eeq
Using the simple form of  $\s_N(x,r_{perp},b_t)\,=\,2 $ in region II,
one can integrate  \eq{DISCRSEC} over $k_2$ and $z$ introducing a new
variable $M^2\,\,=\,\,\frac{k^2_1}{z(1 - z)}$. We obtain
\beq \label{FINQCD} 
\s^{QCD}_{\g^* p}\,\,=\,\,\frac{\alpha_{em} 2 N_c}{3 \pi^2}\sum_{f} Z^2_f
\,\,\int^{\infty}_{M^2_0}\,\,\frac{M^2 d
M^2}{(\,Q^2\,\,+\,\,M^2\,)^2}\,\pi \int^{b^2_0}_0 d b^2_t\,\,.
\eeq
At first sight it appears that there is  no natural cutoff in the above
 $M^2$
integration, but   as we shall   see later this  is not so.
\subsection{The unitarity bound on the  photon cross section}
To obtain the unitarity bound for  the total cross section of the photon -
nucleon interaction, we use the decomposition of \eq{SUMSH} and Gribov's
estimates  for $\s^{soft}$ given in \eq{SOFT}. For $\s_{M^2 N}
(s)$ in \eq{SOFT} we can apply the Froissart - Martin bound of \eq{FRST},
since it is a typical hadronic ( on mass shell ) cross section. 
 Note, that  the Froissart - Martin bound is a high energy limit for
which the Gribov black disc assumption is perfectly adequate since the
diffractive ( $M \,\neq\,M' $ ) channels are  suppressed relative to the
elastic  ( $M\,=\,M' $ ) channel.
  We can evaluate
$\s^{hard}$ in \eq{SUMSH} using the inequality $\s^{hard}\,\leq\,\s^{QCD}$
where $\s^{QCD}$ is determined by \eq{FINQCD}. We wish to explore  further
the nature of  the upper limit of the integration with respect to $M^2$ in
\eq{FINQCD}. To do this we  return to \eq{B0}. 
This equation has no solution if 
\beq \label{LIMM} 
S(b_t = 0)\,\,\frac{ \pi^2 \as}{3}\,r^2_{\perp}\,x
G(x,\frac{4}{r^2_{\perp}})\,\,<\,\,1\,\,.
\eeq
Therefore, the main contribution to  $\s^{QCD}_{\g^* p}$ in \eq{DISCRSEC}
comes from the kinematic region II  with
$r^2_{\perp}\,>\,r^2_0(x)$,
 where $r^2_0 (x) $ is a  solution of the equation
\beq \label{R0}
S(b_t = 0)\,\,\frac{ \pi^2 \as}{3}\,r^2_0(x)\,x
G(x,\frac{4}{r^2_0(x)})\,\,=\,\,1\,\,.
\eeq
Using the notation $S(b_t = 0 ) \,=\,\frac{1}{\pi R^2_N}$ ( see
\eq{BFF}), \eq{R0} can be
rewritten in a more familiar form
\beq \label{R0FIN}
\frac{ \pi \as}{3 \,R^2_N}\,r^2_0(x)\,x 
G(x,\frac{4}{r^2_0(x)})\,\,=\,\,1\,\,.
\eeq
Due to the uncertainty principle
\beq \label{UNCRTNTY}
r^2_{\perp}\,\,\propto\,\,\frac{1}{k^2_{\perp}}\,\,=\,\,\frac{1}{M^2 \, z
( 1 - z )}\,\,.
\eeq
As the integration\footnote{We will comment on the
$z$ 
- integration in the ``hard" cross section for $\Omega\,\ll\,1$
later.}
 over $z$  in \eq{DISCRSEC} is  convergent  in the
limit   $\s_N(x,r_{perp},b_t)\,=\,2 $
( see \eq{MCDSQ} and \eq{FINQCD} ), we can safely put $z = \frac{1}{2}$ in
\eq{UNCRTNTY}\footnote{It means that $r_{\perp}\,\propto\,\frac{2}{M}$.
This fact justifies our main input of a  separation scale
 ( $M_0$ )  in  Gribov's formula of \eq{GF}.}, and rewrite \eq{R0FIN}
as
an
equation for the upper limit of the integration over $M^2$
\beq \label{MX}
\frac{4 \pi \as}{3 \,R^2_N\,M^2(x)}\,\,x
G(x, M^2(x))\,\,=\,\,1\,\,.
\eeq
Collecting all contributions we find
\beq \label{UB}
\s(\g^* N )\,\,\leq\,\,\frac{\alpha_{em}}{3
\,\pi}\,\,\{\,\,C\,\,\ln^2\frac{s}{s_0}\,\,\int^{M^2_0}_{4m^2_{\pi}}
 \,\frac{R(M^2)\,M^2 \,d 
M^2}{(\,Q^2\,+\,M^2\,)^2}\,\,+\,\,2\,R_{\infty}
b^2_0 \,\ln(\,\frac{Q^2 \,+\,M^2(x)}{Q^2\, + \,M^2_0}\,)\,\,\}\,\,.
\eeq
This equation provides  an improved Gribov - Froissart bound for the  photon
- hadron
 total cross section, in place  of the  less restrictive one given in
\eq{GFB}.
For very small
$x$ \eq{MX} leads to $M^2(x)\,\rightarrow\,\Lambda^2 \,\,exp(
\sqrt{a\ln(1/x)})$, where $\Lambda$ is the QCD scale and the constant
$a$ has
been calculated in Ref.\cite{GLR}. This  implies  that for very  small
$x$,
the
  bound for $\s(\g^* N)$ at $Q^2 \,<\,M^2(x)$ is
\beq \label{UBLX}
\s(\g^{*} N )\,\,\leq\,\,\frac{\alpha_{em}}{3
\,\pi}\,\,2\,R_{\infty}\,\,[\, \sqrt{a
\ln(1/x)}\,-\,\ln(Q^2/\Lambda^2)\,]\,\,C\,\ln^2\frac{s}{s_0}\,\,.
\eeq
In the ultra small $x$ limit we obtain that $\s(\gamma^{*}
N)\,\leq\,C'\,( \,\ln(\frac{1}{x}\,)^{\frac{5}{2}}$ where $C'$  contains
all relevant constants.

 \subsection{Numerical estimates for the behaviour of $\s(\g^* N)$}

 We can use \eq{UB}   also  to  make  numerical  estimates of the high
energy
behaviour
of the DIS total cross section. For this purpose we rewrite \eq{UB}  and
attempt to evaluate $\s_{M^2 N}$ of \eq{SOFT} rather than using its high
energy bound. To this end we assume 
\beq \label{UBG}
\s(\g^* N )\,\,\leq\,\,\frac{\alpha_{em}}{3
\,\pi}\,\,\{\,\s^{soft}_{hadron}(s)\,\,\int^{M^2_0}_{4m_{\pi}} 
 \,\frac{R(M^2)\,M^2 \,d 
M^2}{(\,Q^2\,+\,M^2\,)^2}\,\,+\,\,2\,R_{\infty}
b^2_0 \,\ln\frac{Q^2 \,+\,M^2(x)}{Q^2\, + \,M^2_0}\,\,\}\,\,,
\eeq
where $\s^{soft}_{hadron} (s)$ is a typical cross section for a meson -
nucleon interaction. To obtain an estimate, we take  $\s^{soft}_{hadron} 
(s)\,=\,
\frac{1}{2}\,(\,\s(\pi^+ p ) \,+\,\s( \pi^- p)\,)$, and use    the
Donnachie
- Landshoff parameterization \cite{DL} for its energy behaviour.

\begin{figure}[h]
\begin{tabular}{l r}
\psfig{file=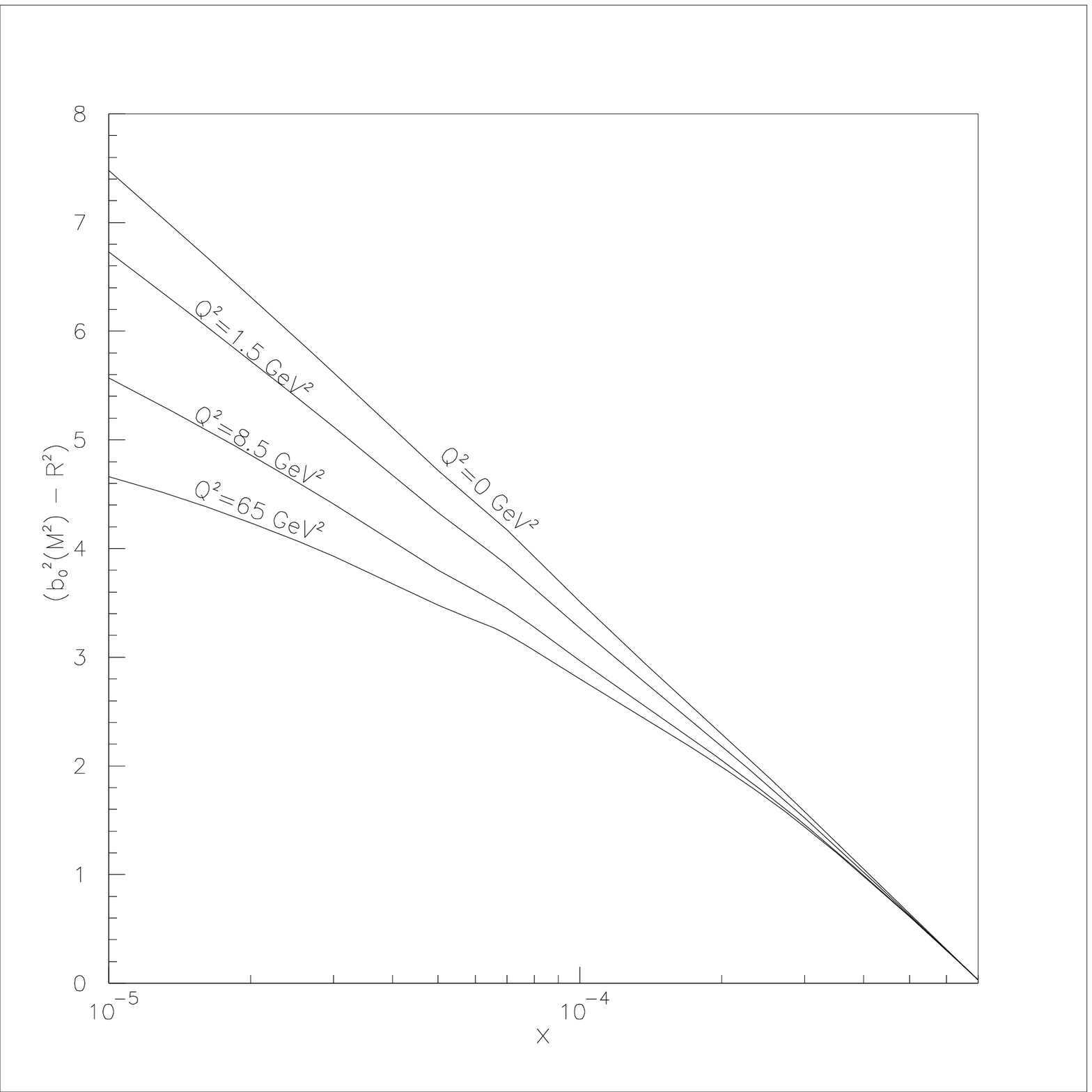,width=70mm} &\psfig{file=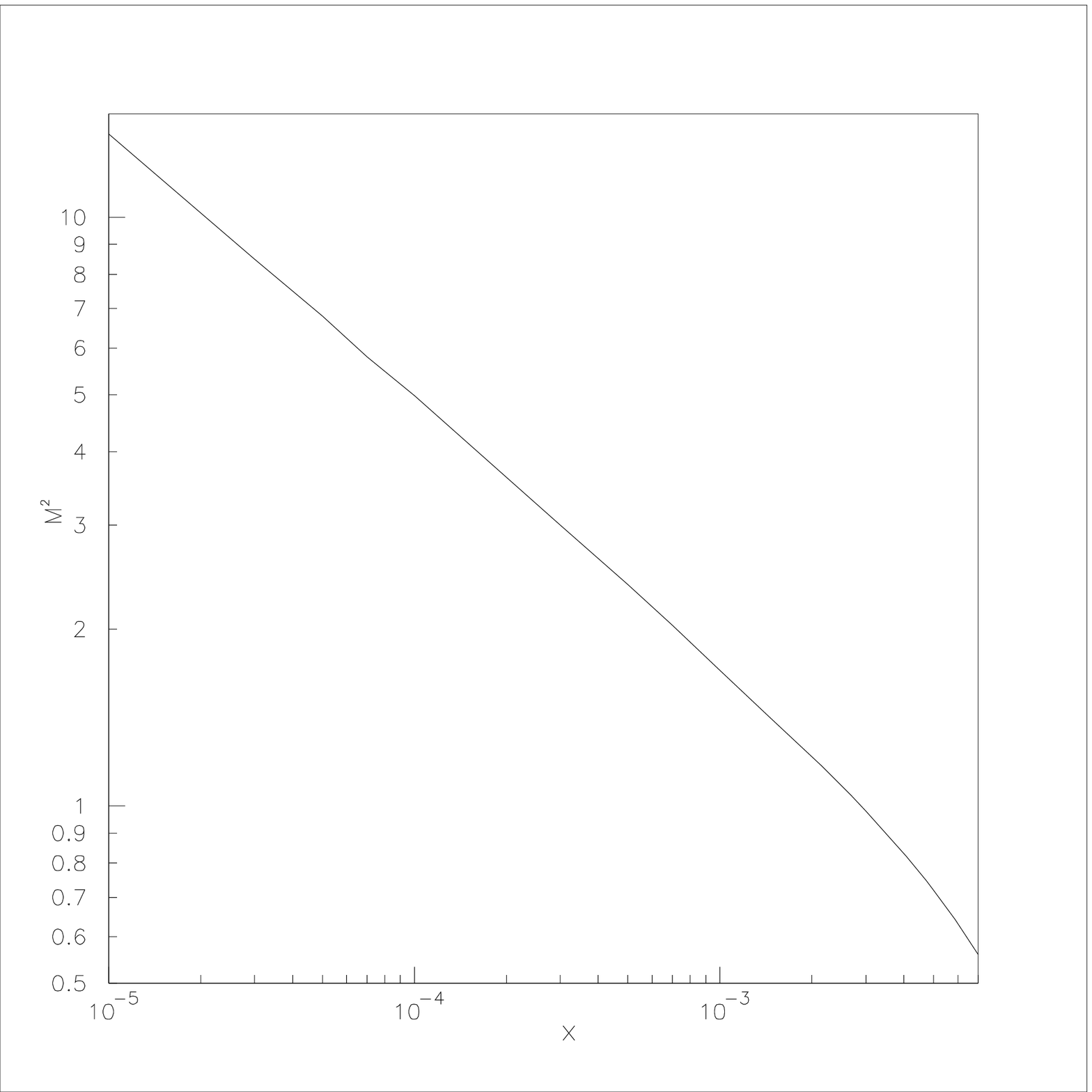,width=70mm}\\
\end{tabular}
\caption{  Solutions of  \protect \eq{B0} ($b^2_0(x)$ ) and of  \protect
\eq{MX} ($M^2(x)$), using the GRV parameterization \protect \cite{GRV} for
the gluon structure function.}
 \end{figure}
Solutions of \eq{B0} ( i.e.  $b^2_0(x)$ ) and of \eq{MX} ( i.e.
$M^2(x)$ ) are
plotted in Fig.2, using the GRV parameterization \cite{GRV} for the gluon
density.
 In  solving \eq{B0}, we have  used  \eq{BFF}, with $R^2_N\,=\,10\,GeV^2$.
 The value of $R^2_N$ is
derived  from HERA
data
on diffractive production of vector mesons in DIS, and  from the  high
energy
phenomenology for ``soft" processes ( see Ref.\cite{GLMSLOPE} for
details ).
 We need to calculate $b^2_0$ at fixed $x$ for all  values of
$M^2\,\leq\,M^2(x)$. However, since  the integral over $M^2$ is
logarithmic,
 we can evaluate  its   contribution at an  average $ {\bar M}^2$. $ {\bar
M}^2$ is determined  from the relation $\int^{{\bar M}^2}_{M^2_0}\frac{d
M^2}{Q^2
\,+\,M^2}$\,=\,$\int^{M^2(x)}_{{\bar M}^2}\frac{d M^2}{Q^2 \,+\,M^2}$,
 which  gives
${\bar
M}^2 = \sqrt{(\,Q^2 \,+\,M^2(x)\,)(\,Q^2\,+\,M^2_0\,)}$\,-\,$Q^2$.
$b^2_0$ at this mass value  is  plotted in Fig.2.
 We consider the values, given  in Fig.2 , to be more relevant
 at presently accessible energies,
than the highly asymptotic Froissart - like estimates.
 
Fig.3 shows the energy dependence of the r.h.s. of \eq{UBG}
together  with the experimental data. The values of $M^2(x)$ ( see Fig.2b )
show that we can trust our estimate, given by \eq{UBG}, only for
$x\,\leq\,10^{-3}$. This  means that we can compare our  bound
only
with the available  experimental data at relatively small values of
$Q^2$. 
 We plot the  data and our estimates from
\eq{UBG}  only at $Q^2 \approx  0$, since for quasi real photoproduction we  
reach
the smallest values of $x$. 
One should note  that
\eq{UBG} was proven only at very small $x$, where we can neglect the
contribution from the kinematic region where $\Omega\,\ll\,1$ ( region I ).
 Actually, in the HERA kinematic region  at all available $Q^2$ and
$W$, we cannot neglect the contribution from region I. On the other hand, 
 Fig.3 shows that our bound is rather close to the experimental data for
real photoproduction. This suggests that \eq{UBG}  reflects the main 
physics for the HERA kinematic region. 
We expect that the inclusion of 
 the ``hard" contribution from kinematic region I,
 will improve our estimates, but will not produce a 
dramatic change.

\begin{figure}
\centerline{\psfig{file=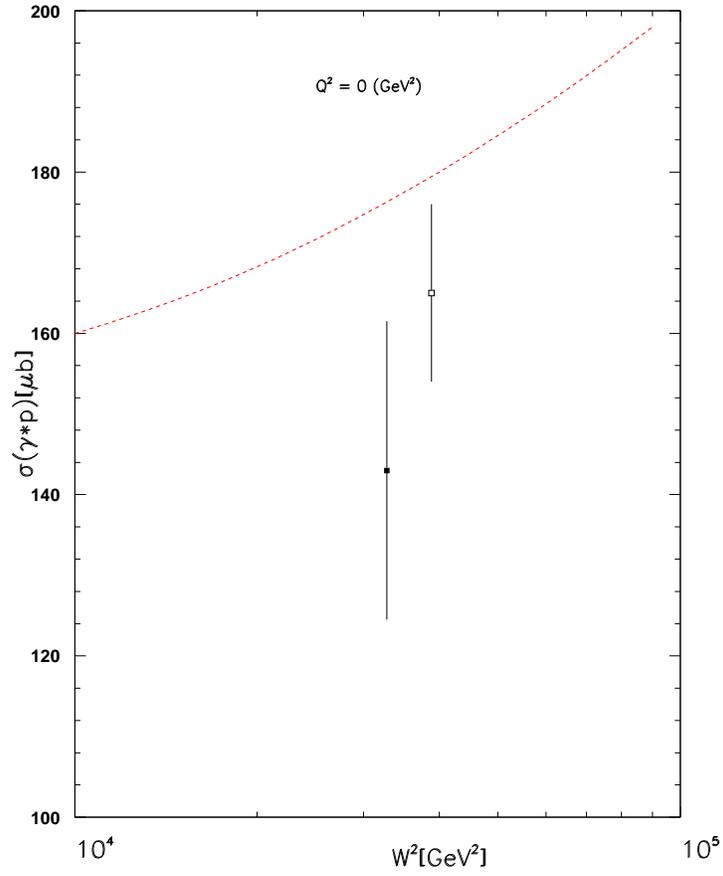,width=100mm}}
\caption{ The energy dependence of the r.h.s. of \protect \eq{UBG}
together 
 with the experimental data.}
\end{figure}

\section{Matching of the  ``soft" and ``hard" processes in DIS}
\subsection{General description}
In the following we  develop a phenomenological approach to describe  DIS
at all
values of $Q^2$ based on the separation of the ``soft" 
and ``hard"  interactions, in the framework of the Gribov
formula ( see \eq{GF} and \eq{SUMSH} ). This  approach provides a
relatively simple description, in which the mass integration with
$M^2\,<\,M^2_0$ is  controlled by the ``soft"
interaction. For $M^2\,>\,M^2_0$ we are dealing with  a ``hard" 
 interaction, which we can treat  in pQCD.
For $\s^{soft}$ in \eq{SUMSH} we assume  that the ``soft" high energy
strong
interaction suppresses  the diffractive dissociation of a hadron state of
mass a $M$ to a hadron state with a  different mass $M'$.  This
property is true, for example,  in the additive constituent quark model, 
where the
interaction of the hadron can be reduced to an interaction of the quarks, 
namely, only the first diagram of Fig.4  ( with
$M = M'$ ), contributes. In a different context this is also a
consequence of the implementation of screening corrections \cite{GLR}.
\begin{figure}
\centerline{\psfig{file=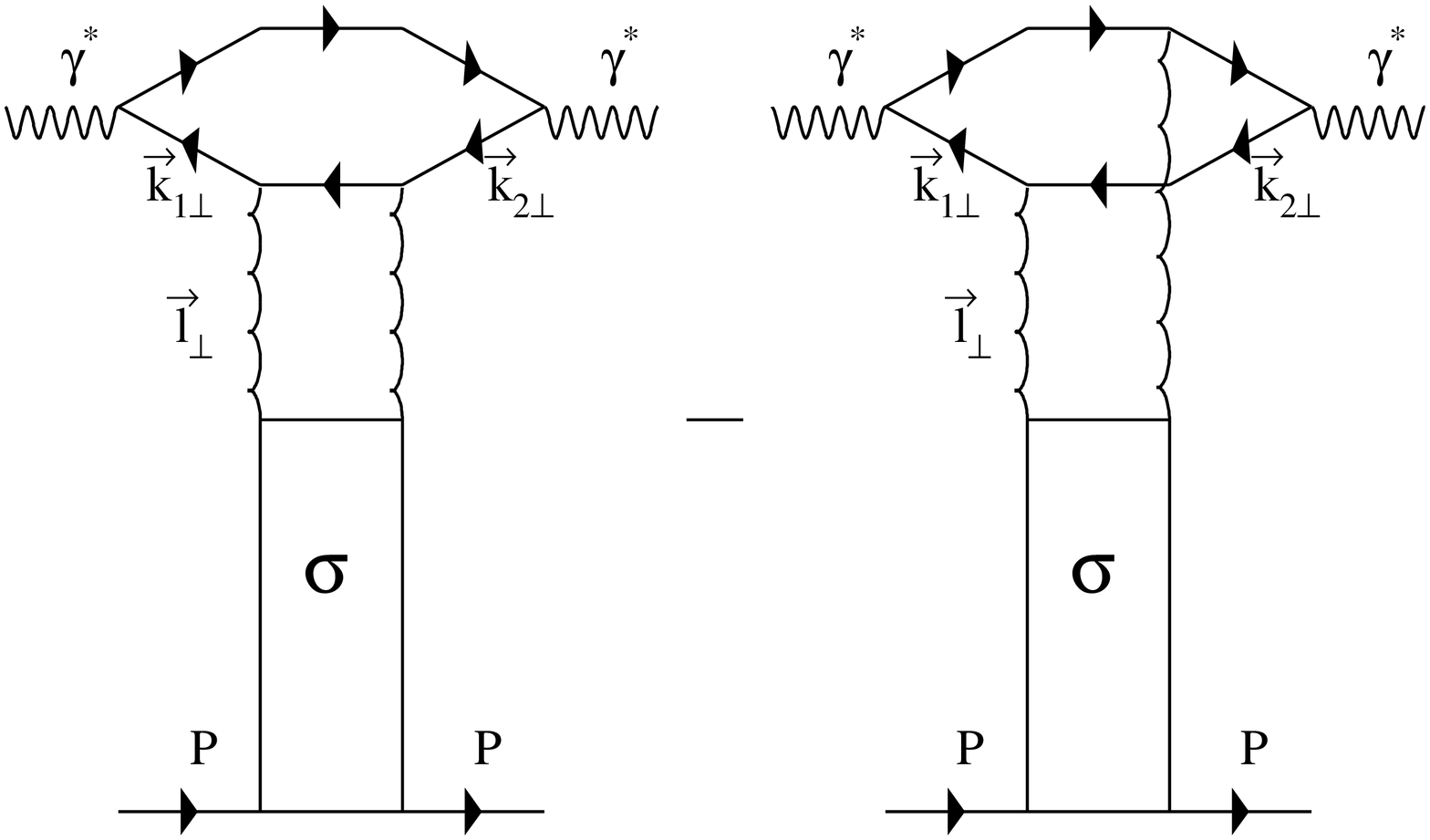,width=100mm}}
\caption{ The diagrams for two gluon exchange model.}
\end{figure}
With this assumption, we can rewrite the ``soft" contribution in the form
\beq \label{SOFTMO}
\s^{soft}\,\,=\,\,\frac{\alpha_{em}}{3 \pi}\,\int^{M^2_0}_{4m_{\pi}}
\,\,\frac{R(M^2)\,M^2\,d M^2}{(\,Q^2\,\,+\,\,M^2\,)^2}\,\,\s_{M^2
N}(s)\,\,.
\eeq 

 For $\s^{hard}$ we use the general formula of \eq{HARD}, where
$\s^{QCD}(M^2,M'^2,s)$ is rewritten in terms of the  gluon - nucleon
interaction in the framework of a two gluon exchange model, shown  in
Fig.4. This model is certainly correct in the region of very small $x$
($\as \ln(1/x)\,\geq\,1$) and large $M^2$ ($\as \ln(M^2/\Lambda^2)
\,\geq\,1$), and  it also  reflects the main properties of the  QCD
interaction outside
this particular kinematic region. 

 We shall discuss the assumptions made for $\s^{soft}$ and \eq{SOFTMO} 
 later. Prior   to that we wish to specify the well known
``hard" contribution. This  will  also be   instructive for our
  discussion of the ``soft" contribution in section 3.3.   
\subsection{The two gluon exchange model}
\eq{DISCRSEC} is
the basic  formula for the two gluon exchange model,
where we use the following representation for $\s(x,r_{\perp})\,=\,\int
d^2b_t \s_N(x,r_{\perp},b_t)$
\beq \label{SIGL}
\s(x,r_{\perp})\,\,=\,\,\int \,d^2 l_{\perp} \,\,\s(l^2_{\perp})\,\,\{
\,\,1\,\,-\,\,e^{i
\vec{l}_{\perp}\cdot\vec{r}_{\perp}}\,\,\}\,\,.
\eeq
One can easily see  that the two terms in \eq{SIGL} just reflect the two
diagrams in Fig.4.
 
Substituting \eq{SIGL} in \eq{DISCRSEC} and using \eq{MCDSQ}, we 
obtain
$$
\s^{hard}\,\,=\,\,\frac{\alpha_{em} N_c}{2 \pi^2}\,\,\sum_{f}\,\,Z^2_f
\,\,
\int^1_0\,d z [\,z^2\,+\,(\,1
\,-\,z\,)^2\,]\,\,\,\,\{\,\int\frac{d^2 k_{1\perp}\,\, k^2_{1 
\perp}}{(\,{\bar
Q}^2\,+\,k^2_{1 \perp}\,)^2}\,\,\int^{\infty}_0\,\s(l^2_{\perp})\,d
l^2_{\perp}\,\,-
$$
\beq \label{SIGL1}
\int\,\frac{d^2 k_{1 \perp}\, d^2 k_{2 \perp}\,\, 
\vec{k}_{1 \perp} \cdot \vec{k}_{2 \perp}}{ (\,{\bar
Q}^2\,+\,k^2_{1 \perp}\,)\,\,(\,{\bar Q}^2\,+\,k^2_{2
\perp}\,)}\,\s(l^2_{\perp} = (\vec{k}_{1 \perp} -
\vec{k}_{2 \perp})^2)\,\,\}
\eeq 
In \eq{SIGL1} we integrate over the angle between $\vec{k}_{1 \perp}$ and 
$\vec{l}_{\perp}$ and introduce a new variable $\tilde M$ 

\beq \label{NEWVAR}
M^2\,\,=\,\,\frac{k^2_{1 \perp}}{z\,( 1 - z)}\,\,;
\,\,\,\,\,\,M'^2\,\,=\,\,\frac{k^2_{2 \perp}}{z\,( 1 -z)}\,\,;\,\,\,
\,\,\,{\tilde M}^2\,\,=\,\,\frac{l^2_{\perp}}{z\,( 1 - z)}\,\,.
\eeq
The physical meaning of $\tilde M$ is clear. Indeed, on the average
$$
\vert\,M^2\,\,-\,\,M'^2\,\vert\,\,=\,\,\vert\,\frac{k^2_{1
\perp}}{z\,(
1 -
z)}\,\,-\,\,\frac{(\,\vec{k}_{1 \perp}\,-\,\vec{l}_{\perp}\,)^2 }{z\,( 1 -
z)}\,\vert \,\,=\,\,\frac{ - 2\,\vec{k}_{1 \perp} \cdot \vec{l}_{\perp}
\,+\,l^2_{\perp}}
{z\,( 1 - z)}\,\,=\,\,< {\tilde M}^2 >.
$$ 
In terms of the new variables, \eq{SIGL1} has  the form
\beq \label{TGF}
\s^{hard}\,\,=\,\,\frac{\alpha_{em} }{4 \pi^2}\,\,
\int^1_0\,d z [\,z^2\,+\,(\,1
\,-\,z\,)^2\,]\,\,\,\,\int\,d l^2_{\perp}\,\int^{\infty}_{M^2_0} 
\,\frac{ R(M^2)\,d
M^2}{Q^2\,+\,M^2}\,\,
\eeq
$$
\{\,\,\frac{M^2\,-\,Q^2}{M^2\,+\,Q^2}\,\,+\,\,\frac{Q^2\,+\,{\tilde 
M}^2\,-\,M^2}
{\sqrt{(\,Q^2\,+\,M^2\,+\,{ \tilde
M}^2\,)^2\,-\,4\,M^2\,{\tilde M}^2}}\,\,\}\,\,
\s(l^2_{\perp})\,\,.
$$
Note that  \eq{TGF} applies   in the region 
$M^2\,>\,M^2_0$ and that  $R(M^2)$ 
replaces
$R_{\infty} \,=\,N_c\,\sum_f\,Z^2_f$.

Since  $z(1-z)\,=\,\frac{l^2_{\perp}}{{\tilde M}^2}$, \eq{TGF}
can be rewritten in the form
$$
\s^{hard}\,\,=\,\,\frac{\alpha_{em} }{4 \pi^2}\,\,
\int^{\infty}_{4m_{\pi}} \,\frac{d {\tilde M}^2}{{\tilde M}^4}\,\,
\int^{\infty}_{M^2_0}\,\,\frac{R(M^2)\,\,d\,M^2}{Q^2\,\,+\,\,M^2}\,\,
\int^{\frac{{\tilde
M}^2}{4}}_{Q^2_0}\,\,[\,1\,-\,2\frac{l^2_{\perp}}{{\tilde
M}^2}\,]\,\,l^2_{\perp}\,d l^2_{\perp}\,\,\s(l^2_{\perp})\,
\frac{1}{\sqrt{1\,-\,\frac{4\,l^2_{\perp}}{{\tilde M}^2}}}
$$
\beq \label{TGFFIN}
\{\,\,\frac{M^2\,-\,Q^2}{M^2\,+\,Q^2}\,\,+\,\,\frac{Q^2\,+\,{\tilde 
M}^2\,-\,M^2}
{\sqrt{(\,Q^2\,+\,M^2\,+\,{\tilde
M}^2\,)^2\,-\,4\,M^2\,{\tilde M}^2}}\,\,\}\,\,.
\eeq
Recalling that
$\s(l^2_{\perp})\,\,=\,\,\as(l^2_{\perp})\frac{\phi(l^2_{\perp})}{l^2_{\perp}}$,
where $
\as(Q^2)\,xG(x,Q^2)\,\,=\,\,\int^{Q^2}\,\,\as(l^2)\,\phi(l^2)\,d l^2$
( see Ref.\cite{GLR} for details), we obtain, in the limit $\frac{4
\,l^2_{\perp}}{{\tilde M}^2}\,\ll\,1$, that
 
\beq \label{TGFINN}
\s^{hard}\,\,=\,\,\frac{\alpha_{em} }{3 \pi}\,\, 2\,\pi^2\,\,
\int^{\infty}_{M^2_0}
\,\frac{d M^2 \,\,R(M^2)}{Q^2\,+\,M^2}\,\,\int^{\infty}_{4 Q^2_0} 
\,\frac{d
{\tilde M}^2}{{\tilde M}^4}
\,\,\as(\frac{{\tilde M}^2}{4})\,x\,G(x,\frac{{\tilde M}^2}{4})
\eeq
$$
\{\,\,\frac{M^2\,-\,Q^2}{M^2\,+\,Q^2}\,\,+\,\,\frac{Q^2\,+\,{\tilde 
M}^2\,-\,M^2}
{\sqrt{(\,Q^2\,+\,M^2\,+\,{\tilde
M}^2\,)^2\,-\,4\,M^2\,{\tilde M}^2}}\,\,\}\,\,.
$$
This equation is our master formula for the evaluation of  the ``hard"
contribution ( with light quarks )
to the
total   photon - nucleon  cross section. $Q^2_0$ is the starting  value of
the gluon
virtuality for the DGLAP evolutuion equations.

We wish to add two comments concerning the above master equation.

1)   $z$ has the same value before and after the 
interaction.  This is a direct manifestation of the leading log
(1/x) approximation in which we only  take into account  contributions
of
the form $ (\,\as \,\ln(1/x)\,)^n $. To understand this we  compare the
time of the interaction  of the  $q \bar q $ - pair   with the
target (
$\tau_i$ ) to 
the life time 
of the virtual photon fluctuating into a
 $q \bar q $ - pair (
$\tau_{\gamma^*}$ ).
 According to the uncertainty principle 
\beq \label{TAUFL}
\tau_{\gamma^*}\,\,\sim\,\,\frac{1}{\Delta E}\,\,=\,\,\vert
\frac{1}{\,q_{-}\,-\,k_{1 -}\,-\,k_{2 -}\,}\vert\,\,=\,\,\frac{z(1 - z)
q_+}{
{\bar Q}^2 \,\,+\,\,k^2_{\perp}}\,\,,
\eeq
where $k_1$ and $k_2$ are the four momenta of quark (antiquark) (see
Fig.4).
An estimate of the interaction time  can be obtained from the typical time
for the emission of a gluon with momentum $l$ from  the quark $k_1$,
 ( i.e., see the second diagram in Fig.4 ). 
\beq \label{TAUINT}
\tau_i\,\,\sim\,\,\vert \frac{1}{k_{1 -} \,\,-\,\,k'_{1 -}\,\,-\,\,l_{-}}
\vert\,\,=\,\,\vert \frac{q_{+}}{\frac{k^2_{1
\perp}}{z}\,\,-\,\,\frac{k^2_{2
\perp}}{z'}\,\,-\,\,\frac{l_{\perp}}{\alpha}}\vert\,\,,
\eeq
where $\alpha\,=\,\frac{l_+}{q_+}$ and $ z'\,=\,z\,-\,\alpha$. In the
leading log(1/x) approximation we have $ \alpha\,\ll\,z$ and hence
\beq \label{TAU}
\tau_i\,\,\approx\,\,\frac{\alpha
q_+}{l^2_{\perp}}\,\,\ll\,\,\tau_{\gamma^*}\,\,.
\eeq 
 \eq{TAU} shows explicitly that the processes of a virtual
and/or  real photon scattering off a target can be described as a two
stage process: initially the photon decays into a $q \bar q$ - pair and
only a long time after that  the pair interacts with the target. This is
why such
a process can be described by
 the Gribov's formula.

2) An  important feature of \eq{TGFINN} is the fact that we
integrate
over $z$ using $z ( 1 - z) \,=\,\frac{l^2_{\perp}}{{\tilde M}^2}$ . The
integral over $l^2_{\perp}$ is  logarithmic  and the typical values
of $l^2_{\perp}$ are  of the order of $ l^2_{\perp}\,\,\approx\,\,{\tilde
M}^2 exp( - \frac{1}{\gamma})$ where $\gamma$ is the anomalous dimension.
Since we are interested  in the region of small $x$ where
$\gamma$ is rather large, one can see that the  typical values of $z$,
which
are
essential in our integration, turn out to be of the order of unity.
As a result the contribution of the aligned jet model ( AJM )
\cite{AJM} is small at least at high energies.   Actually, an explanation
showing that the
emission of many gluons suppresses the nonperturbative  AJM - type
configurations  has been given  by Gribov and Lipatov \cite{DGLAP} more
than two decade ago. They showed that the emission
of  multi gluons generates a  Sudakov form factor which suppresses the
AJM - like  contributions.
Note  that the result of the $z$ integration is very important
for  the understanding of our approach since it justifies our idea that
$M^2$ is a
good measure of  the typical distances involved in the process. Indeed,
 $r^2_{\perp}\,\propto\,\frac{1}{M^2}$ holds only for  $z \,\sim\,1$ ( see
\eq{UNCRTNTY} ). These estimates have been made  for the total
cross section. For exclusive channels the situation is quite different.
For example, for the diffractive dissociation in DIS, induced by a
transverse polarized photon, the typical distances are rather large   and
do
not depend on the value of the  mass in the Gribov's formula.
 
 For  heavy quarks, the diagrams of Fig.4 give 

\beq \label{TGHQ}
\s^{hard}_{\bar Q Q}\,\,=\,\,\frac{\alpha_{em} }{3 \pi}\,\,
2\,\pi^2\,\,
\int^{\infty}_{4 m^2_Q}
\,\frac{dM^2 \,\,R^{QQ}(M^2)}{Q^2\,+\,M^2}\,\,\int^{\infty}_{ Q^2_0} 
\,\frac{d
{\tilde M}^2}{{\tilde M}^4}
\,\,\as(\frac{{\tilde M}^2}{4})\,x\,G(x,\frac{{\tilde M}^2}{4})
\eeq
$$
\{\,\,\frac{M^2\,-\,Q^2}{M^2\,+\,Q^2}\,\,+\,\,\frac{Q^2\,+\,{\tilde 
M}^2\,-\,M^2}
{\sqrt{(\,Q^2\,+\,M^2\,+\,{\tilde 
M}^2\,)^2\,-\,4\,(\,M^2\,-\,4m^2_Q\,)\,{\tilde 
M}^2}}\,\,\}
\,\,,     
$$                                                                        
where $R^{QQ}$ is the heavy quark contribution to the ratio in \eq{RATIO},
and $m_Q$ is the mass of the heavy quark. In \eq{TGHQ} we have  assumed  that
the
quark is heavy enough to be described in pQCD without any contribution of
 the ``soft" processes.
In the  above formulae  $x \,\,=\,\,\frac{Q^2\,+\,M^2}{W^2}$, and  $W$ is
the
energy of the photon - nucleon interaction.
\subsection{A model for  the ``soft"  interactions}
Our model for the ``soft" interactions is based on \eq{SOFTMO}.
We observe  that our discussion on  the time structure
of the photon - hadron interaction does not  depend on any
specific properties of QCD, and can be considered, therefore, as a  
main feature of the
parton model approach to  high energy photon induced  interactions ( see
Ref.
\cite{FEYN}
 ). It means that, the approximation  $z \,=\,z'$ in Fig.1 applies also 
for the ``soft"
interactions of our model. 

In the parton model \cite{FEYN},  as well as in the high energy
phenomenology
\cite{DL},  the ``soft" processes at high energy can be described by the 
exchange of a soft  Pomeron which has the property of   Regge
factorization \cite{COLLINS}
$$
\s_P (s, M,M')\,\,=\,\,g_P(M,M')G_P (\,\frac{s}{M^2}\,)^{\alpha_P(0)
\,-\,1}\,\,,
$$
where $\alpha_P(t)$ is the Pomeron trajectory and $g_P( M,M')$ and  $G_P$
are vertices of the Pomeron interaction with the  quark - antiquark pair
and 
with the  proton respectively.  Substituting this equation in
\eq{DISCRSEC} one
can see that only the photon wave function depends on $z$. This
integral
is convergent  and  $z$ is typically about unity also  for a  photon with
large
virtuality. Coming back to \eq{NEWVAR} one can see that $M'$ could be
different from $M$ only if $k_{2 \perp}\,\gg \,k_{1 \perp}$ ( or $k_{2
\perp}\,\ll \,k_{1 \perp}$ ). On the other hand, in the parton model the
typical transverse momenta (  $l_{\perp} $ in Fig.4 ) are of the order
of the ``soft" scale, i.e. about  $1\, GeV$. 
Therefore, in a photon - hadron interaction we expect that the value of
$M'$
cannot be much larger than the value of $M$. The  success of the
additive quark model ( AQM ) ( see the first diagram in Fig.4 )  in the
description
of the high energy scattering (
see Ref. \cite{DL} )  supports our  assumption that the
difference ( $ M' - M$ )  is much smaller than the transverse momentum
scale
 for the Pomeron. 

We can, therefore, rewrite the general Gribov formula of \eq{GF} in the
form
\newpage
$$
\s(\gamma^* N)\,\,=\,\,
$$
$$
\frac{\alpha_{em}}{3
\pi}\,\,\int^{M^2_0}_{4m_{\pi}}\,\,\frac{R(M^2) M^2  d
M^2}{(\,Q^2\,\,+\,\,M^2\,)^2}\s_{M^2N}(s)\,\,\int^{M}_{- M }\,\frac{ M d
\Delta M}{M^2}
 \frac{g_P(M,M')}{g_P(M,M)}\,\frac{M^2 \,\,+\,\,Q^2}{ (M + \Delta M)^2
\,\,+\,\,Q^2}
$$
\beq \label{SOFTGF}
\,\,=\,\,\,\,\frac{\alpha_{em}}{3
\pi}\,\,\int^{M^2_0}_{4m_{\pi}}\,\,\frac{R(M^2) M^2  d
M^2}{(\,Q^2\,\,+\,\,M^2\,)^2}{\tilde\s}_{M^2N}(s) \,\,,
\eeq
where $\Delta M \,\,=\,\,M' \,-\,M$.

Note that the $\Delta M$ integration is particular to photon induced
reactions and is, obviously absent in the case of hadron - hadron 
collisions. This correction, as well as other relevant corrections, are
absorbed in our definition  of ${\tilde \s}_{M^2 n}$ in \eq{SOFTGF}.
\newline
In light of  the above discussion, we choose  \eq{SOFTMO} and /or
\eq{SOFTGF} as the master formula for our description of the ``soft"
contribution.

The  quantities appearing in  \eq{SOFTMO} which  need to be  specified  
are 
the ratio
$R(M^2)$, and  the interaction cross
section   of a hadronic system with mass $M$  with the target  ( $\s_{M^2 
N}(s)$ in \eq{SOFTMO} ).
 Although,  there is  experimental data for $R(M^2)$ ( see
Ref. \cite{EEEXP} ),
 we have used the parameterization of Ref.\cite{R} for $ R(M^2)$
 \footnote{We thank E. Gurvich for drawing our attention to
Ref.\cite{R}}. The two main
ingredients of
this parameterization, reproducing the   experimental data,  are the
resonance
contribution and the  background, which  at high masses approaches  the
QCD
result ( $R(M^2)\,\rightarrow \ R_{\infty}$ ).
We approximate ${\tilde\sigma}_{M^2 N}$  in the resonance region by
 \beq \label{RCRSEC}
\s_R\,\,=\,\,\kappa\, \frac{1}{2}\,(\,\s(\pi^ + p)\,\,+\,\,\s(\pi^- +
p)\,)
\eeq
where we use   the Donnachie - Landshoff  Reggeon parameterization
\cite{DL} for the
cross
section
of pion - proton interaction, 
\beq \label{DL}
\s_R\,\,=\,\,\kappa\,[\,A \, \, ( \frac{s}{s_0})^{
\alpha_P(0)\,-\,1}\,\,+\,\,B\,\, (
\frac{s}{s_0})^{\alpha_R(0)\,-\,1}\,]\,\,,
\eeq
with a Pomeron and Reggeon trajectory intercepts of  $
\alpha_P(0)$\,=\,1.079
and $\alpha_R(0)$\,=\,0.55 ; $s_0$\,=\,1 \,$GeV^2$; A\,=\,13.7 mb and
B\,=\,31.9 mb. We introduce a rescaling  constant  $\kappa$ as a parameter
in
\eq{RCRSEC} and
\eq{DL}.
 $\kappa$ = 1 corresponds to a simplified AQM where  the
integrant is taken  at 
  $\Delta
M $ = 0  and first order corrections of $\frac{\Delta M}{M}$
and
$\frac{ \Delta M}{m_G}$, where $m_G$ is the typical scale of the soft
Pomeron, are neglected. This is  discussed further in section 3.4.

 To describe the interaction of the  background contribution in $R(M^2)$
, we need to determine  the correct energy variable for the
interaction of the hadronic system of  mass $M$ with the target. We
suggest to  replace  the variable $ \frac{s}{s_0} $ by $ \frac{1}{x_M}$,
where $x_M\,=\,\frac{M^2}{s}$. In other words, we replace \eq{DL} by 
\beq \label{DLM}
{\tilde\s}_{M^2 N}\,\,=\,\,\kappa\,[\,A' \, \, ( \frac{1}{x_M})^{
\alpha_P(0)\,-\,1}\,\,+\,\,B'\,\, (
\frac{1}{x_M})^{\alpha_R(0)\,-\,1}\,]\,\,,
\eeq
where $A'$ = 13.1 mb and $B'$ = 41.08 mb. The values of $A'$ and $B'$ were 
chosen
so that  \eq{DL} for the  $\rho$   meson - proton  interaction is valid.
The
variable
$x_M$ is not
unique, but it correctly  describes  the high energy interaction,
in all
parton - like models in the region of large  mass $M$ ( see
Ref.\cite{GLR}
for example ).    
   
\subsection{ Comparison with the experimental data}
To compare with the  experimental data, we calculate the  total cross
section of
the photon - proton interaction using the following formula
\beq \label{MODEL}
\s( \g* p )\,\,=\,\,\s^{soft} (\, \eq{SOFTMO},\,\,\eq{DLM}\, )\,\,
+\,\,\s^{hard}_{\bar
q q} (\,\eq{TGFINN}\,)\,\,+\,\,\s^{hard}_{\bar Q Q}(\,\eq{TGHQ}\,)\,\,.
\eeq

\eq{MODEL} depends on two free parameters: $\kappa$ and $M^2_0$ and the
input gluon distribution $x G(x,\frac{{\tilde M}^2}{4} )$. In Fig.5 we
plot $\s(\g^* p )$ as a function of  $W^2$ for various values of $Q^2$ and
compare
with the relevant experimental data. In our calculations we have used
the GRV parameterization \cite{GRV} in higher order of pQCD (GRVHO). We
have chosen the parameters   $\kappa$ = 0.6 and $M^2_0$ = 5 $GeV^2$ so as
to obtain a
reasonable  reproduction of the data.  

As can be seen from Fig.5, we obtain a good fit fot the low $Q^2$ data
over the  entire $W$ range. 
However, at $Q^2$ higher than  a few $GeV^2$, our description of the data
is
deficient in as much as we are not in agreement with the low energy
experimental points and our predicted  high energy behaviour is steeper
than the
data. Below we elaborate on these features as well as on some important
details of the suggested model.

\begin{figure}
\centerline{\psfig{file=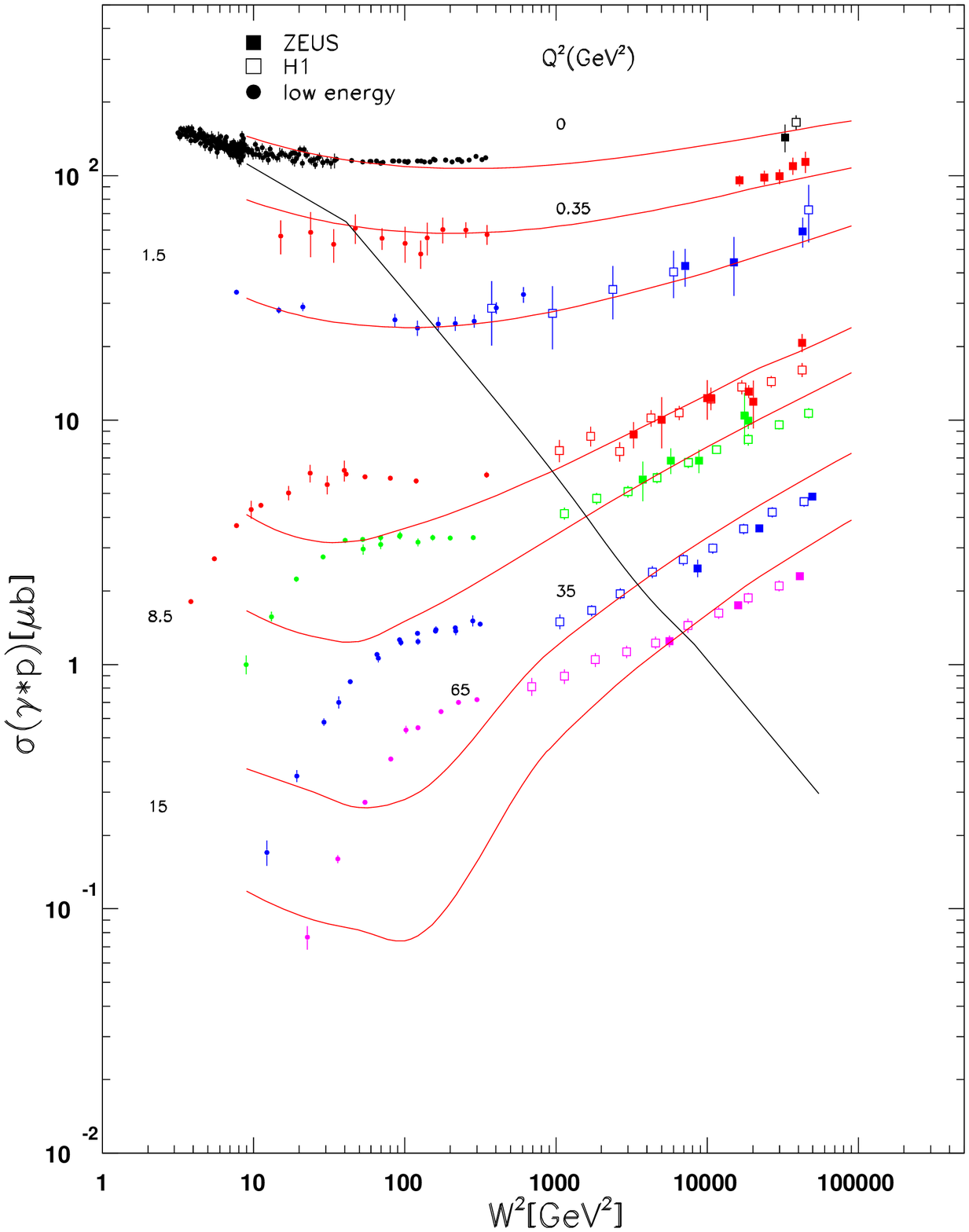,width=160mm}}
\caption{ The comparison of the experiment data ( see
Ref.\protect\cite{HERADATA}
and references therein) for $\sigma( \gamma^* p
)$ and our calculation, using \protect\eq{MODEL}. The diagonal line
indicates the boundary for $x = 10^{-2}$.}
\end{figure}

1. With only two free parameters, we manage to reproduce the energy
dependence for real photoproduction and DIS with  $Q^2\,<\,8  GeV^2$ 
cross
sections.
 This is shown in Fig.6 where we compare the recent high energy -
low $Q^2$ ZEUS  data \cite{ZEUSDATA} with the predictions of our model.
Our
results compare  favorably  with the Donnachie-Landshoff\cite{DL} and
GRV \cite{GRV} parameterizations.

\begin{figure}
\centerline{\psfig{file=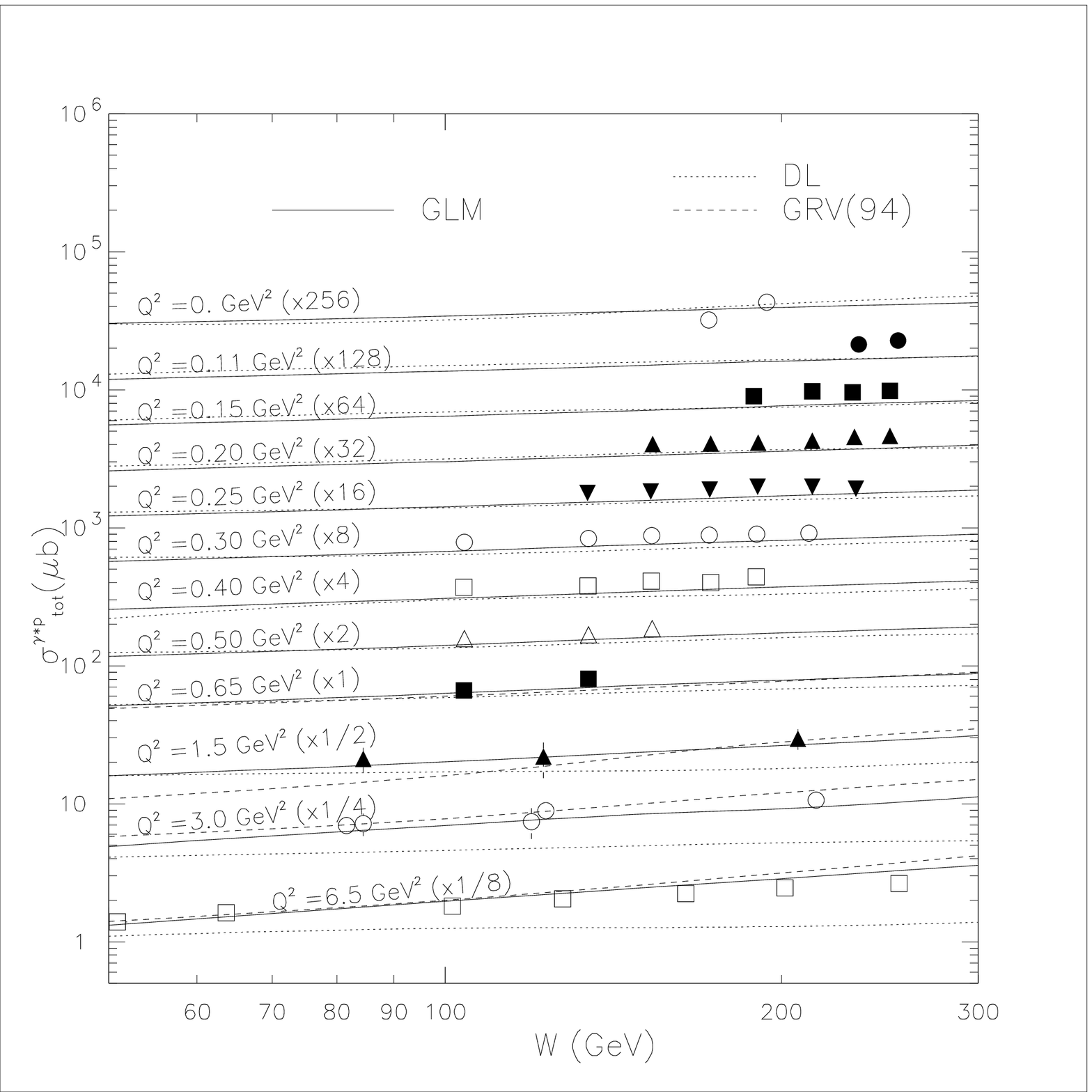,width=160mm}}
\caption{ Low $Q^2$ and high $W$  data  from ZEUS ( see
Ref.\protect\cite{ZEUSDATA} ), compared to our predictions ( solid lines),
Donnachie - Landshoff approach\protect\cite{DL} ( dotted lines ) and the
GRV
parameterization \protect\cite{GRV} ( dashed lines).}
\end{figure}

2. The fact that we are unable  to  reproduce  the low energy
behaviour of the higher
$Q^2$  ( $x \,>\,10^{-2}$ ) data is not surprizing. These data correspond
to higher $x$ values
for which  the two gluon approach to DIS is insufficient. In Fig.5 we show
 the line corresponding $x$ = 0.01 which illustrates this
point. Clearly for high  $x$ ( low $W$ ) one should also include the
contribution coming from 
the $Q^2$  evolution of the quark distribution as a part of the pQCD
description of  DIS. We discuss below how to expand our formalism so as to
include this input as well. The fit  can also  be
improved by the utilization of other  input parton distributions which are
less
steep than GRVHO in the small $x$ limit.

3.  An unexpected  feature of our results  is that we require a   
value of $\kappa\,$ = 0.6 to rescale the AQM  estimate of the ``soft"
contribution. As we have noted, this value of $\kappa$ reflects  the need
to integrate over $\Delta M$ which is particular to photoproduction and
DIS and does not appear in hadron - hadron scattering, where the AQM has
been 
checked experimentally.  Our  result  is different from  VDM
 \cite{VDM}, where in order to describe
the  
experimental data,  one has  to assume that the vector meson -  nucleon  
cross
section is  bigger than the  AQM estimates. This difference arises  from
the  background contribution that is neglected in the VDM approach.
It should be stressed that such a contribution, which is included in our
parameterization of $R(M^2)$,  is needed to reproduce the
$Q^2$ - dependence, which is  much smoother than the VDM 
prediction.    In addition to uncertainties from the $\Delta M $
integration we also see at least two reasons
leading to a value of $\kappa$ smaller than unity.
Our evaluation suggests  that  using the approximate formula of Ref.
\cite{R} we overestimate the experimental data  by about 10\%
 ( $\kappa\,\approx$\,0.9 from this source ).  The second
source  is the need for shadowing corrections
(SC ).
  SC definitely diminish the value  of the cross section. We can  estimate
 the SC  from the  value of the
diffractive dissociation cross section,  using the AGK cutting rules    
\cite{AGK},
 which  relates the SC to the total cross
section,  namely, $\Delta
\s^{SC}\,=\,\s^{DD}$, where $ \s\,=\,\s^{AQM}\,-\,\Delta \s^{SC}$.
 The experimental value of the diffractive cross section is about 15\%,   
both from  real photoproduction and from   pion - proton interaction.
These two sources   suggest  a  value of $\kappa
\,\approx\,0.7$ to which an additional 15\% reduction is added due to the
$\Delta M $ integration.

4. The inclusion of heavy quark contribution is  important.   In Fig.7 we
plot the ratio
 $R^{CHARM}_{LIGHT}\,\,=\,\,\frac{\s ( charm\,\,\,quarks)}
{\s(light\,\,\,quarks)}$
as a function of energy at different values of $Q^{2}$.  This
ratio is rather small for real photoproduction, reaching 5\% at high
energies. For large values of $Q^2$ the ratio approaches 30 -  35\% 
at  high energies  $W \,\geq\,30 \,GeV$.

\begin{figure}[h]
\centerline{\psfig{file=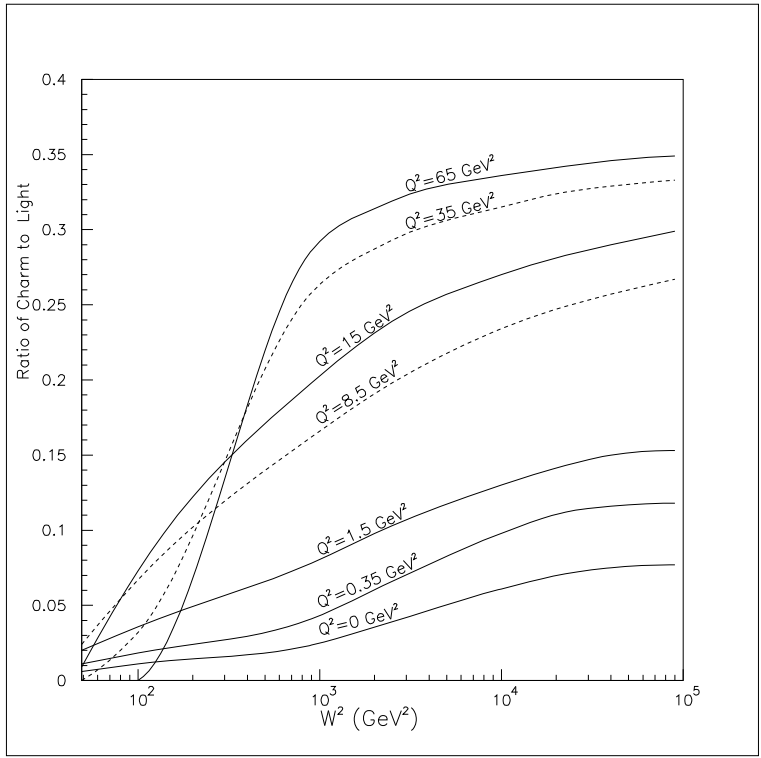,width=100mm}}
\caption{The ratio   $R^{CHARM}_{LIGHT}\,\,=\,\,\frac{\s
(charm\,\,\,quarks)}
{\s(light\,\,\,quarks)}$  as function of $W^2$ for different values of
$Q^2$ .}
\end{figure}

In general, our model provides a very simple approach to incorporate the
``soft" and ``hard" components of photoproduction and DIS. In  the 
detailed fit of the data we observe two
features that we consider to be rather general.

1.  The high energy  ``hard" contribution turns out to be sizable, even
for $Q^2$ = 0. To illustrate, how important the ``hard" contribution is
in
our formalism, we plot in Fig.8  the ratio
$R^H_S\,=\,\frac{\s^{hard}(\g^* p)}{\s^{soft}(\g^* p)}$.
One can see that for $Q^2 = 0$,  $R^H_S\,\approx\,1\% $ at W \,=\,10 \,GeV
and it grows
to $R^H_S\,\approx\,15\% $ at W = 300 \,GeV ( $s$ = $W^2$ ). This increase
is sufficiently large,  that
it alone may  account for  the experimentally observed increase in
the  energy dependence of the total
cross section, for real photoproduction. In other words, it is possible to 
fit the
experimental data assuming that the ``soft" Pomeron ( see \eq{DL} ) has an  
intercept which is equal to  unity ( $\alpha_P(0)$\,=\,1 ).

Our result suggests    a possible and  interesting interpretation for  the
origin
of the experimentally observed increase of the total cross sections for
hadron -
hadron interactions. As for each hadron we also have a contribution
 of the ``hard" process to the total cross section, due to  the
probability that   two quarks approach each other   at a sufficiently
small distance. The probability for this  ``hard"  process is controlled
by
the respective wave functions of the interacting hadrons.

\begin{figure}[h]
\begin{tabular}{l r}
\psfig{file=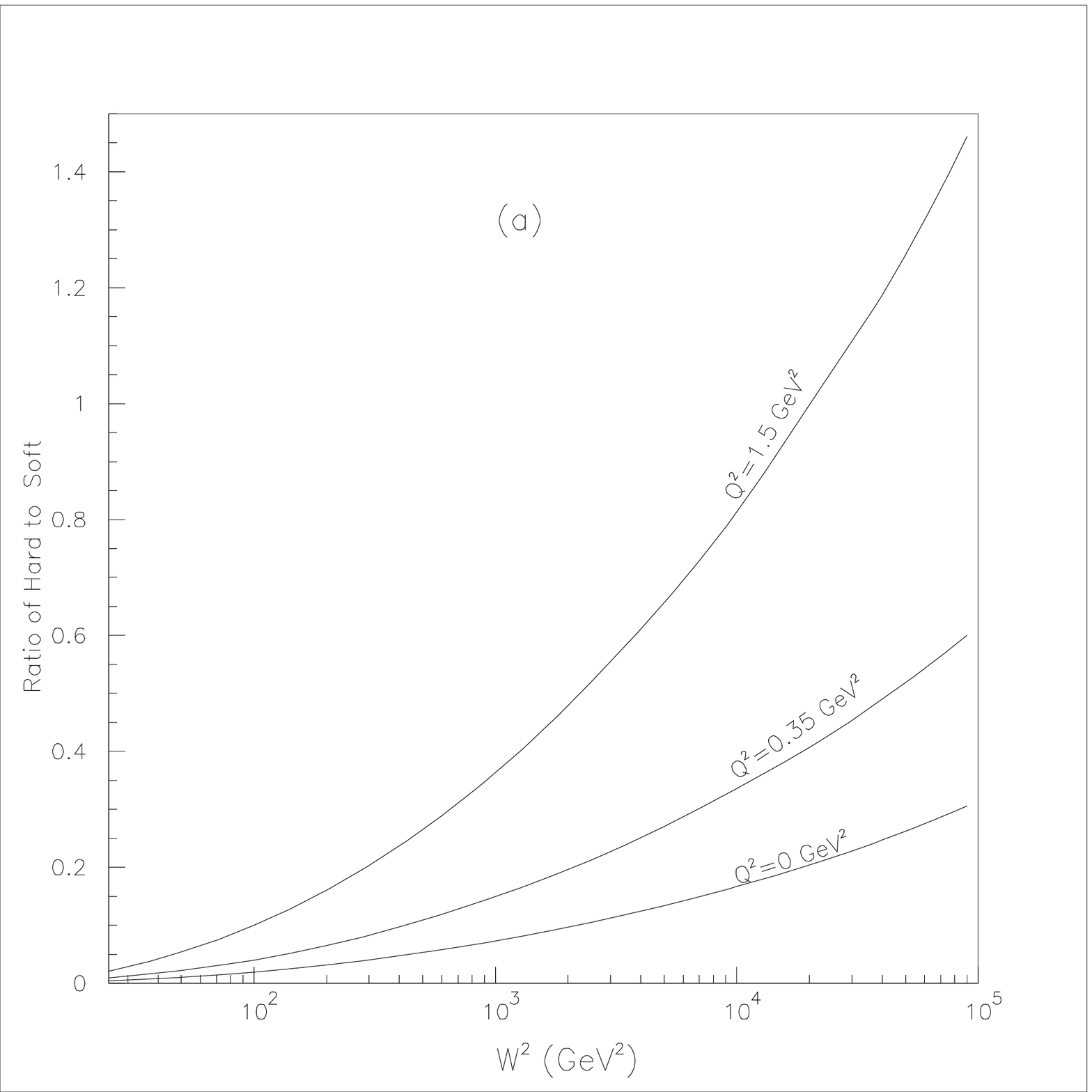,width= 80mm} &
\psfig{file=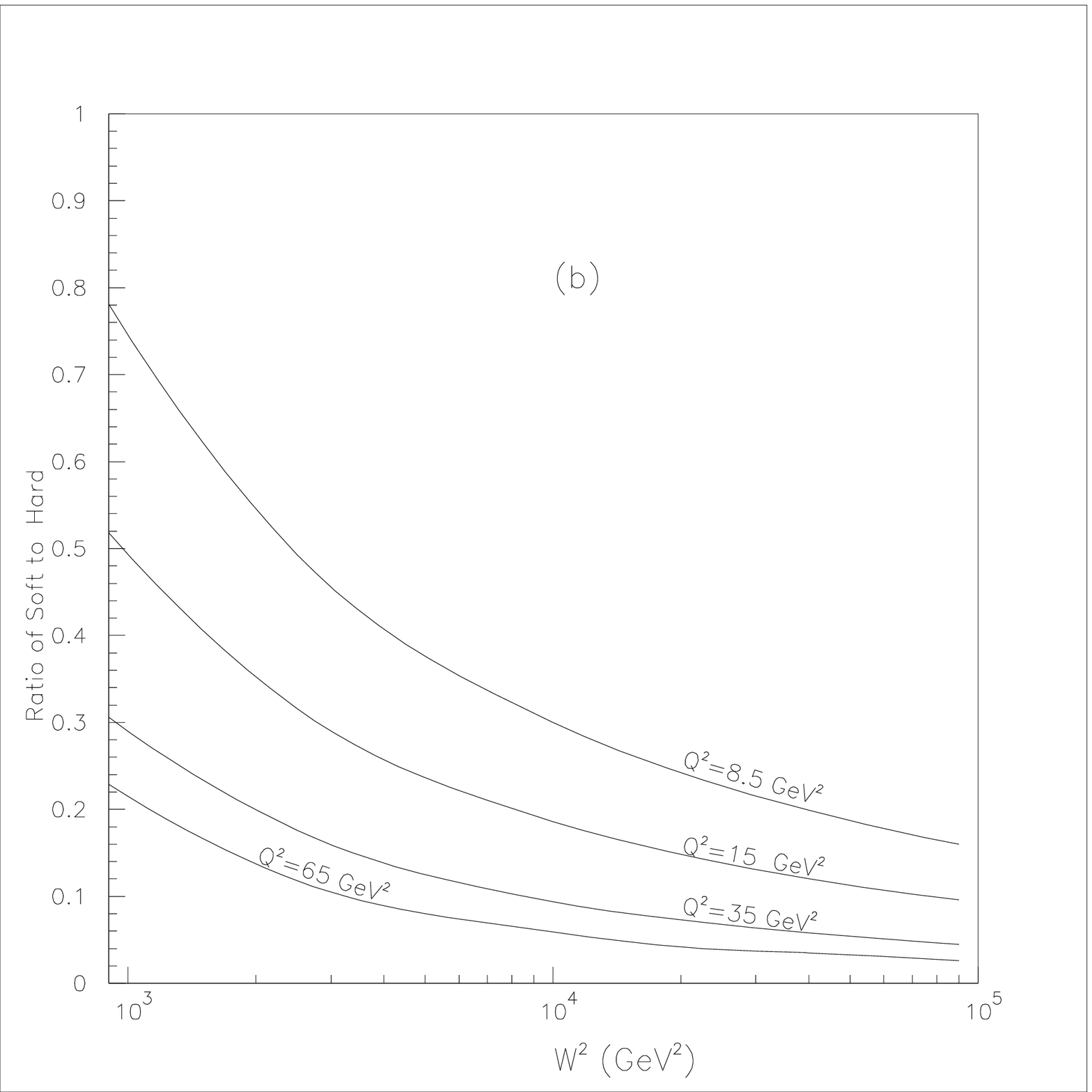,width= 80mm} \\
\end{tabular}
\caption{The ratios (a)  $R^H_S\,=\,\frac{\s^{hard}(\g^*p)}{\s^{soft}(\g^*
p)}$  and (b)  $R^S_H\,=\,\frac{\s^{soft}(\g^* p)}{\s^{hard}(\g^*p)}$ 
   as function of $W^2$ for different values of $Q^2$ .}
\end{figure}

2. We find  contamination of the ``hard" processes by the ``soft" ones.
For example, at $Q^2\,=\,15 \,GeV^2$ which is usually
considered   a large value for $Q^2$, the ratio
$R^S_H\,=\,\frac{\s^{soft}(\g^* p)}{\s^{hard}(\g^* p)}$ changes from
 1  at W = 30 GeV to 0.2 at W = 300 GeV.  Even at $Q^2\,=\,65\, GeV^2$,
\,\,$R^S_H$ is about 0.12 at W = 100 GeV.
 The above results, lead one to view
DIS in a new light,  and provide a basis for a better
 understanding of what is meant by   small distances or high
photon virtualities.

\subsection{A generalization of our model to  larger $x$}
 The  generalization is based on \eq{TGFINN}, utilizing the fact that for
the DGLAP \cite{DGLAP} evolution equation the following equation holds in
the region
of small $x$ ( see \cite{EKL} for example )
\beq \label{F2G}
\frac{\partial F^{DGLAP}_2}{\partial \ln ( Q^2/\Lambda^2)}\,\,=\,\,
\frac{2 \as}{9\pi}x G^{DGLAP}(x,Q^2)\,\,.
\eeq
We  suggest  replacing  $\as x G(x, Q^2)$ in \eq{TGFINN} by
$\frac{\partial F^{DGLAP}_2}{\partial \ln ( Q^2/\Lambda^2)}$ and
 use this generalized formula for DIS, even in the region of $x$  not very
small. Note, that for $x$  not too small, we obtain
\newline
 ( see Ref.
\cite{EKL}
for example)  a more general
formula for
$\frac{\partial F^{DGLAP}_2}{\partial \ln ( Q^2/\Lambda^2)}$ than \eq{F2G}
 which includes the quark densities. After doing so, \eq{TGFINN}
reduces to the form
\beq \label{GF2}
\s^{hard}\,\,=\,\,3\,\pi^2\,
\alpha_{em}\,\,\int^{\infty}_{M^2_0}\,\frac{R(M^2)\,d M^2}{Q^2\,+\,M^2}
\,\,\int^{\infty}_0 \,\,\frac{d {\tilde
M}^2}{{\tilde M}^2}\,\,\frac{\partial
F^{DGLAP}_2 (x,\frac{{\tilde M}^2}{4})}{\partial {\tilde M}^2}
\eeq
$$   
\{\,\,\frac{M^2\,-\,Q^2}{M^2\,+\,Q^2}\,\,+\,\,\frac{Q^2\,+\,{\tilde 
M}^2\,-\,M^2}
{\sqrt{(\,Q^2\,+\,M^2\,+\,{\tilde
M}^2\,)^2\,-\,4\,M^2\,{\tilde M}^2}}\,\,\}\,\,.
$$
Integrating \eq{GF2} by parts one  obtains
\beq \label{GF2FIN}
\s^{hard}\,\,=\,\,3\,\pi^2\,
\alpha_{em}\,\,\int^{\infty}_{M^2_0}\,\frac{R(M^2)\,d M^2}{Q^2\,+\,M^2}
\,\,\int^{\infty}_0 \,\,\frac{d {\tilde M}^2}{{\tilde M}^4}\,\,F^{DGLAP}_2
(x,\frac{{\tilde M}^2}{4})
\eeq
$$
\{\,\,\frac{M^2\,-\,Q^2}{M^2\,+\,Q^2}\,\,+\,\,\frac{Q^2\,+\,{\tilde 
M}^2\,-\,M^2}
{\sqrt{(\,Q^2\,+\,M^2\,+\,{\tilde
M}^2\,)^2\,-\,4\,M^2\,{\tilde M}^2}}\,\,-\,\,
\frac{4\,Q^2\,M^2\,{\tilde
M}^2}{[\,\sqrt{(\,Q^2\,+\,M^2\,+\,{\tilde M}^2\,)^2\,-
\,4\,M^2\,{\tilde M}^2}\,]^3}\,\,\}\,\,.
$$
Actually, a  formula of the same  type  as  \eq{GF2FIN}  was first 
suggested by
Badelek and Kwiecinski \cite{BK}, but using our formalism we obtain quite
a different result. We can consider \eq{GF2FIN} as a
generalization of the Badelek - Kwiecinski approach. In addition to the
resonances we have also included  the background contribution, and obtain
the
 contributions of both ``soft" and ``hard" processes  by
integrating over
$M^2$ and ${\tilde M}^2$ in \eq{GF2FIN}. 

Numerical results pertaining to \eq{GF2FIN} will be published separately.
\section{Conclusions}
 We have  achieved two goals in this  paper:
 
1. We provide an explanation of  how and why the short  distance   
 ( ``hard" ) interaction,
calculable in pQCD, provides a  mass cutoff in the Gribov's
formula for  photon - hadron collisions. We have shown that the
Gribov  bound ( see \eq{GFB} ) given in Ref.\cite{AFS}),  overestimates
the photon - hadron cross section, and should be replaced by a more
restrictive limit , as derived in this paper ( see \eq{UB},\eq{UBLX} and
\eq{UBG} ).
At fixed $Q^2$ as $W \,\rightarrow \,\infty$ our bound is  $\s(\g^*
p)\,\leq\,
C' ( \ln \frac{1}{x} )^{\frac{5}{2}}$.

2. We developed a simple model which consists of two contributions: 
``soft"  and ``hard".
The ``soft" term describes  the long distance  contribution,
 while the ``hard" term is
 related only to the  short  distance interaction controlled by
pQCD
( and  the DGLAP evolution equation \cite{DGLAP} ). This simple 
 model with only two  parameters  provides a good description of
the available experimental data  over a wide range of  $W$ and $Q^2
\,<\,8\,
GeV^2$. We have suggested  a technique of how to improve the high
$Q^2$
results at sufficiently small values of energy $W$.

Examining our model we found  two interesting features that may be
more general:

a)  Short   distance   effects contribute 
 even at $Q^2\,=\,0$  for  high energies. The contribution is
sufficient  to
explain the energy rise of the total cross section for photoproduction,
which has  been interpreted as an  argument that the 
``soft"
Pomeron has an  intercept larger  than 1 ( see Ref.\cite{DL} ).
This result encourages us to reconsider this  widely  held  explanation,
and to 
estimate the contributions of the ``hard" processes  to the
 growth of
hadron - hadron cross sections, with increasing energy.

b) The long  distance  processes contribute to the total cross
section even at rather large values of $Q^2$. For example, at $Q^2 = 65 \,
GeV^2$ and  $W = 100 \,  GeV$ they are responsible for 10\% of the total 
cross
section. This observation can be very important for understanding  the
energy dependence, as well as the value of the cross sections of other
processes such as  diffractive dissociation, inclusive production 
etc. We propose  to examine these processes  in the
near  future using the same approach.

 Our approach is not in contradiction with the usual description of 
``hard" processes, based on the DGLAP evolution equations with initial
 nonperturbative parton densities  at $Q^2$ = $Q^2_0$. However, we
significally enlarged   the region of applicability of such an approach,
noting that the quark - antiquark pair with large mass can be treated in
pQCD even at $Q^2 = 0$. It allowed us to separate the nonperturbative
contribution in a different way than usually done  and to calculate  a
part of the
initial parton densities at $Q^2 = Q^2_0$ in pQCD.

In general, the  model suggested  allows one  to
discuss the interface between long  and short  distance processes,
  not only on the  qualitative level but also 
on a quantitative
one.  We are of the opinion that our model incorporates what is
known, both theoretically and phenomenologically about ``soft" and ``hard"
physics, and  provides  a method  to estimate the different 
contributions to a
variety of processes. It also allows one  to specify the kinematic
region where the ``hard"
contribution  dominates, and  to calculate it within the framework of 
pQCD.

{\bf Acknowledgements:} We would like to thank our
 colleagues  at Tel Aviv,
E. Gurvich for providing us with a program to evaluate $R(M^2)$ and A.
Levy for making  his file containing the experimental data on
$\s(\g^* p)$ available to us. E.G. and E.L. would like to acknowledge the 
kind 
hospitality of the Theory Group at DESY where this paper was completed.
U.M. thanks LAFEX-CBPF (Rio de Janeiro ) for their hospitality and suport. 
This research was partially supported by THE ISRAEL SCIENCE FOUNDATION
founded by the Israel Academy of Sciences and Humanities.

\end{document}